\documentclass{sig-alternate}

\usepackage{graphicx,subfigure,url}

\renewcommand{\leq}{\leqslant}
\renewcommand{\geq}{\geqslant}

\begin{document}

\conferenceinfo{SIGCOMM'06,} {September 11--15, 2006, Pisa,
Italy.}
 \CopyrightYear{2006}
\crdata{1-59593-308-5/06/0009}

\title{Systematic Topology Analysis and Generation \\ Using Degree Correlations}

\author{
    Priya Mahadevan \\ UC San Diego  \and
    Dmitri Krioukov \\ CAIDA  \and
    Kevin Fall \\ Intel Research  \and
    Amin Vahdat \\ UC San Diego \and
     {\large \{pmahadevan,vahdat\}@cs.ucsd.edu,
     dima@caida.org, kevin.fall@intel.com}
}
\date{}
\maketitle

\begin{abstract}

Researchers have proposed a variety of metrics to measure
important graph properties, for instance, in social, biological, and
computer networks.  Values for a particular graph metric may capture a
graph's resilience to failure or its routing efficiency.
Knowledge of
appropriate metric values may influence the engineering of future
topologies, repair strategies in the face of failure, and
understanding of fundamental properties of existing
networks. Unfortunately, there are typically no algorithms to generate
graphs matching one or more proposed metrics and there is little
understanding of the relationships among individual metrics or their
applicability to different settings.

We present a new, systematic approach for analyzing network
topologies.  We first introduce the $dK$-series of probability distributions
specifying all degree correlations within $d$-sized subgraphs of a
given graph~$G$.  Increasing values of~$d$ capture progressively more
properties of~$G$ at the cost of more complex representation of the
probability distribution.
Using this series, we can quantitatively measure the distance between
two graphs and construct random graphs that accurately reproduce virtually
all metrics proposed in the literature.  The nature of the $dK$-series
implies that it will also capture any future metrics that may be proposed.
Using our approach, we construct graphs
for~\mbox{$d=0,1,2,3$} and demonstrate that these graphs reproduce,
with increasing accuracy, important properties of measured and modeled
Internet topologies. We find that the~\mbox{$d=2$} case is sufficient
for most practical purposes, while~\mbox{$d=3$} essentially
reconstructs the Internet AS- and router-level topologies exactly.  We
hope that a {\em systematic\/} method to analyze and synthesize
topologies offers a significant improvement to the set of tools
available to network topology and protocol researchers.

\end{abstract}


\section{Introduction}
\label{sec:intro}
Knowledge of network topology is crucial for understanding and
predicting the performance, robustness, and scalability of network
protocols and applications.  Routing and searching in networks,
robustness to random network failures and targeted attacks, the speed of
worms spreading, and common strategies for traffic engineering and
network management all depend on the topological characteristics of a
given network.

\begin{figure}[t]
    \centerline{
        \includegraphics[width=2.9in]{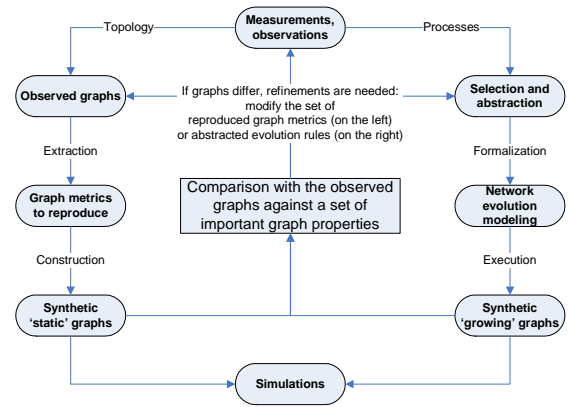}
        \label{fig:static-growing}
    }
    \caption{\footnotesize Methodologies of network topology research.}
\end{figure}

Research involving network topology, particularly Internet topology,
generally investigates the following questions:

\begin{enumerate}

\item \underline{generation}: can we efficiently generate ensembles of random but ``realistic'' topologies by reproducing a set of simple graph metrics?

\item \underline{simulations}: how does some (new) protocol or application perform on a set of these ``realistic'' topologies?

\item \underline{evolution}: what are the forces driving the evolution (growth) of a given network?

\end{enumerate}

Figure~\ref{fig:static-growing} illustrates
the methodologies used to answer these questions
in its left, bottom, and right parts, respectively.
Common to all of the methodologies
is a set of practically-important graph properties used
for analyzing and comparing sets of graphs at the center box of the figure.
Many such properties have been defined and explored in the literature.
We briefly discuss some of them in Section~\ref{sec:metrics}.
Unfortunately, there are no known algorithms to construct random graphs with
given values of most of these properties, since they typically
characterize the global structure of the topology, making it difficult
or impossible to algorithmically reproduce them.

This paper introduces a finite set of reproducible graph properties, the
{\em $dK$-series}, to describe and constrain random graphs
in successively finer detail. In the limit, these properties describe
any given graph completely.
In our approach, we make use of probability distributions,
the {\em $dK$-distributions}, specifying node degree correlations
within subgraphs of size~$d$ in some given input graph. We call {\em $dK$-graphs\/}
the sets of graphs constrained by given values of $dK$-distributions. Producing a family of
$0K$-graphs for a given input graph requires reproducing only the {\em
average\/} node degree of the original graph, while producing a
family of $1K$-graphs requires reproducing the original graph's node
degree distribution, the $1K$-distribution. $2K$-graphs reproduce the joint
degree distribution, the $2K$-distribution, of the original graph---the probability
that a randomly selected link connects nodes of degrees~$k$ and~$k'$.
$3K$-graphs consider interconnectivity among triples of nodes, and so forth.  Generally, the set of
\mbox{$(d+1)K$}-graphs is a subset of $dK$-graphs.  In other words, larger values
of~$d$ further constrain the number of possible graphs. Overall, larger values of
$d$ capture increasingly complex properties of the original graph.
However, generating $dK$-graphs
for large values of $d$ also become increasingly computationally complex.

A key contribution of this paper is to define the series of $dK$-graphs and
$dK$-distributions and to employ them for generating and analyzing
network topologies. Specifically, we develop and implement
new algorithms for constructing $2K$- and $3K$-graphs---algorithms to generate
$0K$- and $1K$-graphs are already known.
For a variety of measured and modeled Internet AS- and router-level topologies,
we find that reproducing their $3K$-distributions
is sufficient to accurately reproduce {\em all\/} graph properties we have
encountered so far.

Our initial experiments suggest that the $dK$-series has the potential
to deliver two primary benefits. First, it can serve as a basis for
classification and unification of a variety of
graph metrics proposed in the literature. Second, it establishes a path
towards construction of random graphs matching any complex graph properties,
beyond the simple per-node properties considered by existing approaches
to network topology generation.

\section{Important Topology Metrics}
\label{sec:metrics}
In this section we outline a list of graph metrics that
have been found important in the networking literature. This list is
not complete, but we believe it is sufficiently diverse and
comprehensive to be used as a good indicator of graph similarity in
subsequent sections. In addition, our primary concern is how accurately
we can reproduce {\em important\/} metrics.
One can find statistical analysis of these metrics for Internet
topologies in~\cite{VaPaVe02a} and, more recently, in~\cite{MaKrFo06}.

The {\em spectrum\/} of a graph is the set of eigenvalues of its
Laplacian~$\mathcal{L}$. The matrix elements of~$\mathcal{L}$
are $\mathcal{L}_{ij}=-1/(k_{i}k_{j})^{1/2}$
if there is a link between a $k_{i}$-degree node~$i$
and a $k_{j}$-degree node~$j$; otherwise they are~$0$, or $1$ if $i=j$.
All the eigenvalues lie between 0 and 2. Of particular
importance are the smallest non-zero and largest eigenvalues,
$\lambda_{1}$ and~$\lambda_{n-1}$, where~$n$ is the graph size.
These eigenvalues provide tight bounds for a number of critical network
characteristics~\cite{chung97} including
{\em network resilience\/}~\cite{TaGoJaShWi02} and
{\em network performance\/}~\cite{LiAlWiDo04}, i.e., the
maximum traffic throughput of the network.

The {\em distance distribution}~\mbox{$d(x)$} is the number
of pairs of nodes at a distance~$x$, divided by the total number
of pairs~$n^2$ (self-pairs included). This metric is a normalized
version of {\em expansion\/}~\cite{TaGoJaShWi02}. It is also
important for evaluating the performance of routing
algorithms~\cite{KrFaYa04} as well as of the speed with which worms
spread in a network.

{\em Betweenness\/} is the most commonly used measure of
centrality, i.e., topological importance, both for nodes and links.
It is a weighted sum of the number of shortest paths passing through
a given node or link. As such, it estimates the potential traffic
load on a node or link, assuming uniformly distributed traffic
following shortest paths. Metrics such as {\em link value\/}~\cite{TaGoJaShWi02} or
{\em router utilization\/}~\cite{LiAlWiDo04} are directly related to betweenness.

Perhaps the most widely known graph property is the {\em node degree distribution}~$P(k)$, which specifies the probability of nodes having degree $k$ in a graph. The unexpected finding in~\cite{FaFaFa99} that degree distributions in Internet topologies closely follow power
laws stimulated further interest in topology research.

The {\em likelihood}~$S$~\cite{LiAlWiDo04} is the sum of products of
degrees of adjacent nodes. It is linearly related to the {\em
assortativity coefficient}~$r$~\cite{newman02} suggested as a
summary statistic of node interconnectivity: assortative
(disassortative) networks are those where nodes with similar
(dissimilar) degrees tend to be tightly interconnected. They are more
(less) robust to both random and targeted removals of nodes and links.
Li {\it et al.} use~$S$
in~\cite{LiAlWiDo04} as a measure of graph randomness
to show that router-level topologies are not ``very random'':
instead, they are the result of sophisticated engineering design.

{\em Clustering}~$C(k)$ is a measure of how close neighbors of the
average $k$-degree node are to forming a clique: $C(k)$ is the ratio of
the average number of links between the neighbors of $k$-degree nodes
to the maximum number of such links~${k \choose 2}$.
If two neighbors of a node
are connected, then these three nodes form a triangle
(3-cycle). Therefore, by definition, $C(k)$ is the average
number of 3-cycles involving $k$-degree nodes.
Bu and Towsley~\cite{BuTo02} employ clustering to estimate
accuracy of topology generators. More recently, Fraigniaud~\cite{fraigniaud05}
finds that a wide class of searching/routing strategies are more efficient
on strongly clustered networks.

\section{{\it\rm\large\lowercase{d}K}-series and {\it\rm\large\lowercase{d}K}-graphs}
\label{sec:degcor}
There are several problems with the graph metrics in the previous
section. First, they derive from a wide range of studies, and no
one has established a systematic way to determine which metrics
should be used in a given scenario.  Second, there are no known
algorithms capable of constructing graphs with desired values for
most of the described metrics, save degree distribution and more
recently, clustering~\cite{SeBo05}.
Metrics such as spectrum, distance distribution, and betweenness characterize
global graph structure, while known approaches to generating graphs
deal only with local, per-node statistics, such as the degree
distribution.
Third, this list of metrics is
incomplete.  In particular, it cannot include any future metrics that
may be of interest. Identifying such a metric might result in
finding that known synthetic graphs do not match  this new metric's
value: moving along the loops in
Figure~\ref{fig:static-growing} can thus continue forever.

To address these problems, we focus on establishing a finite set
of mutually related properties that can form a basis for any topological
graph study. More precisely, for any graph~$G$, we wish to identify a
{\em series\/} of graph properties $\mathcal{P}_d$,
\mbox{$d=0,1,\ldots$}, satisfying the following requirements:

\begin{enumerate}
\item {\em constructibility}: we can construct graphs having these properties;

\item {\em inclusion}: any property~$\mathcal{P}_d$ subsumes all
properties~$\mathcal{P}_i$ with~\mbox{$i=0,\ldots,d-1$}:
that is, a graph having property~$\mathcal{P}_d$ is guaranteed to also have
all properties~$\mathcal{P}_i$ for~\mbox{$i<d$};

\item {\em convergence}: as~$d$ increases, the set of graphs having
property~$\mathcal{P}_d$ ``converges'' to~$G$: that is, there exists a
value of index~$d=D$ such that all graphs having
property~$\mathcal{P}_D$ are isomorphic to~$G$.
\end{enumerate}

\begin{figure}
        \includegraphics[width=2.8in]{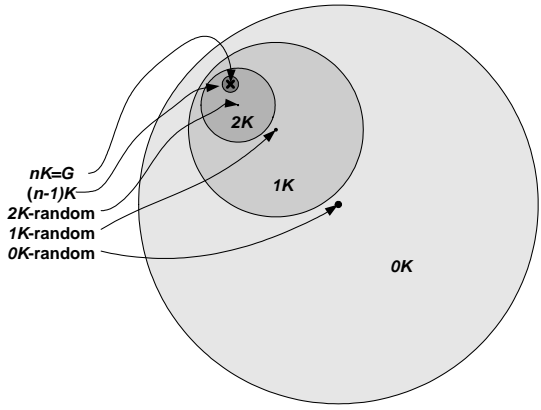}
        \caption{The $dK$- and $dK$-random graph hierarchy.
        {\normalfont The circles represent $dK$-graphs, whereas their centers
        represent $dK$-random graphs. The cross is the $nK$-graph
        isomorphic to a given graph~$G$.}}
        \label{fig:dk}
\end{figure}

In the rest of this section, we establish our construction of the
properties~$\mathcal{P}_d$, which we will call the {\em
$dK$-series}.
We begin with the observation that the most basic properties of a
network topology characterize its connectivity.
The coarsest connectivity property is the {\em average node
degree}~\mbox{$\bar{k}=2m/n$}, where~\mbox{$n=|V|$} and~\mbox{$m=|E|$}
are the numbers of nodes and links in a given graph~$G(V,E)$.  Therefore,
the first property~$\mathcal{P}_0$ in our $dK$-series~$\mathcal{P}_d$
is that the graph's average degree~$\bar{k}$ has the same value as in
the given graph~$G$.
In Figure~\ref{fig:dk} we schematically depict the set of all graphs
having property~$\mathcal{P}_0$ as $0K$-graphs, defining the largest circle.
Generalizing, we adopt the term {\em $dK$-graphs\/} to represent the
set of all graphs having property~$\mathcal{P}_d$.

The $\mathcal{P}_0$ property tells us the average number of links per
node, but it does not tell us the distribution of degrees across
nodes.  In particular, we do not know the number of nodes~$n(k)$ of each
degree~$k$ in the graph.  We define property $\mathcal{P}_1$ to
capture this information: $\mathcal{P}_1$ is therefore the property
that the graph's {\em node
degree distribution}~\mbox{$P(k)=n(k)/n$}\footnote{Sacrificing a certain
amount of rigor, we interchangeably use the enumeration of
nodes having some property in a given graph, e.g., $n(k)/n$, with the
probability that a node has this property in a graph ensemble, e.g., $P(k)$.
The two become identical when $n \to \infty$;
see~\cite{BoPaVe04} for further details.} has the same form as in
the given graph~$G$. It is convenient to call~$P(k)$ the {\em
$1K$-distribution}.
$\mathcal{P}_1$ implies at least as much information
about the network as~$\mathcal{P}_0$, but not vice versa:
given~$P(k)$, we find~\mbox{$\bar{k}=\sum kP(k)$}.
$\mathcal{P}_1$ provides more information than~$\mathcal{P}_0$, and it is
therefore a more restrictive metric: the set of $1K$-graphs
is a subset of the set of $0K$-graphs.
Figure~\ref{fig:dk} illustrates this inclusive relationship
by drawing the set of $1K$-graphs inside the set of $0K$-graphs.

Continuing to~$d=2$, we note that the degree
distribution constrains the number of nodes of each degree in the
network, but it does not describe the interconnectivity of nodes with
given degrees.  That is, it does not provide any information on the
total number~$m(k,k')$ of links between nodes of degree~$k$ and~$k'$.
We define the third property $\mathcal{P}_2$ in our series as the
property that the graph's {\em joint degree distribution}~(JDD) has the same
form as in the given graph~$G$. The JDD, or the {\em $2K$-distribution}, is
\mbox{$P(k_1,k_2)=m(k_1,k_2)\mu(k_1,k_2)/(2m)$}, where $\mu(k_1,k_2)$
is~2 if \mbox{$k_1=k_2$} and~1 otherwise.  The JDD
describes degree correlations for {\em pairs} of connected nodes.
Given $P(k_1,k_2)$, we can calculate $P(k)=(\bar{k}/k)\sum_{k'}P(k,k')$, but not vice versa.  Consequently, the set
of $2K$-graphs is a subset of the $1K$-graphs. Therefore,
Figure~\ref{fig:dk} depicts the smaller $2K$-graph circle
inside~$1K$.

We can continue to increase the amount of connectivity information
by considering degree correlations among greater numbers of
connected nodes.  To move beyond $2K$, we must begin to distinguish
the various
geometries that are possible in interconnecting $d$~nodes.  To
introduce~$\mathcal{P}_3$, we require the following two components:
1)~{\em wedges}: chains of 3~nodes connected by 2~edges, called
the $P_\wedge(k_1,k_2,k_3)$ component; and 2)~{\em triangles}:
cliques of 3~nodes, called the $P_\triangle(k_1,k_2,k_3)$
component:\\
\centerline{\includegraphics[width=2in]{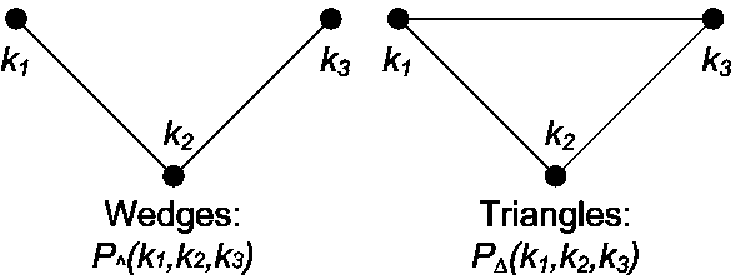}}
As the two geometries occur with different frequencies among nodes
having different degrees, we require a separate probability
distribution for each configuration. We call
these two components taken together the {\em $3K$-distribution}.

\centerline{\includegraphics[width=2.75in]{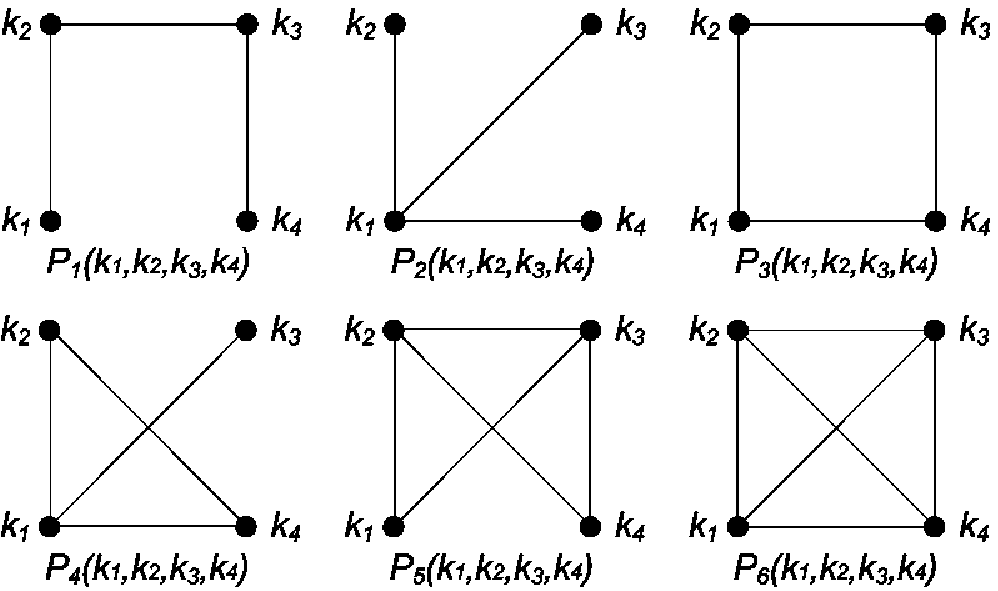}}
For~$\mathcal{P}_4$, we need the above six distributions: where
instead of indices~\mbox{$\wedge,\triangle$} we use
for~\mbox{$d=3$}, we have all non-isomorphic graphs of size~$4$
numbered by~\mbox{$1,\ldots,6$}. We note that the order of
$k$-arguments generally matters, although we can permute any pair
of arguments corresponding to pairs of nodes whose swapping leaves
the graph isomorphic. For example: $P_\wedge(k_1,k_2,k_3) \neq
P_\wedge(k_2,k_1,k_3) \neq P_\wedge(k_1,k_3,k_2)$, but
$P_\wedge(k_1,k_2,k_3) = P_\wedge(k_3,k_2,k_1)$.

In the following figure, we illustrate properties~$\mathcal{P}_d$,
$d=0,\ldots,4$, calculated for a given graph~$G$ of size~$4$,
where for simplicity, values of all distributions~$P$ are the total
numbers of corresponding subgraphs, i.e., \mbox{$P(2,3)=2$} means that
$G$~contains $2$~edges between $2$- and $3$-degree nodes.
\centerline{\includegraphics[width=2.5in]{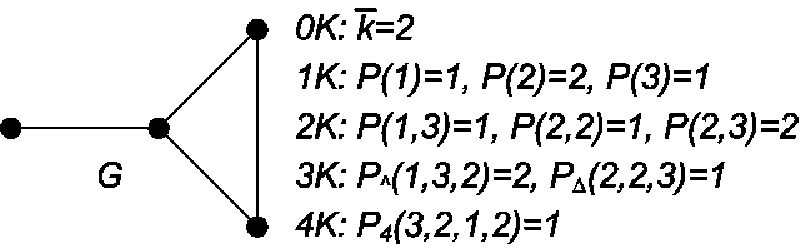}}

Generalizing, {\em we define the {\em $dK$-distributions} to be degree
correlations within non-isomorphic simple connected
subgra\-phs of size~$d$ and the {\em $dK$-series}~$\mathcal{P}_d$ to be the series of
properties constraining the graph's $dK$-distribution to the same form
as in a given graph~$G$.}  In other words, $\mathcal{P}_d$ tells us how groups of
$d$-nodes with degrees~\mbox{$k_1,...,k_d$} interconnect. In
the~`$dK$' acronym, `$K$' represents the standard notation for node degrees,
while `$d$' refers to the number of \textit{d}egree arguments~$k$ of the
$dK$-distributions~$P(k_1,\ldots,k_d)$
and to the upper bound of the \textit{d}istance between nodes with specified degree correlations.
Moving from~$\mathcal{P}_d$ to~$\mathcal{P}_{d+1}$ in describing a given
graph~$G$ is somewhat similar to including the additional
\mbox{$d+1$}'th~term of the Fourier (time) or Taylor series representing a
given function~$F$. In both cases, we describe wider ``neighborhoods''
in~$G$ or~$F$ to achieve a more accurate representation of the
original structure.

\begin{figure*}[tbh]
    \centerline{
        \subfigure[$0K$-graph]
        {\includegraphics[width=1.55in,angle=90]{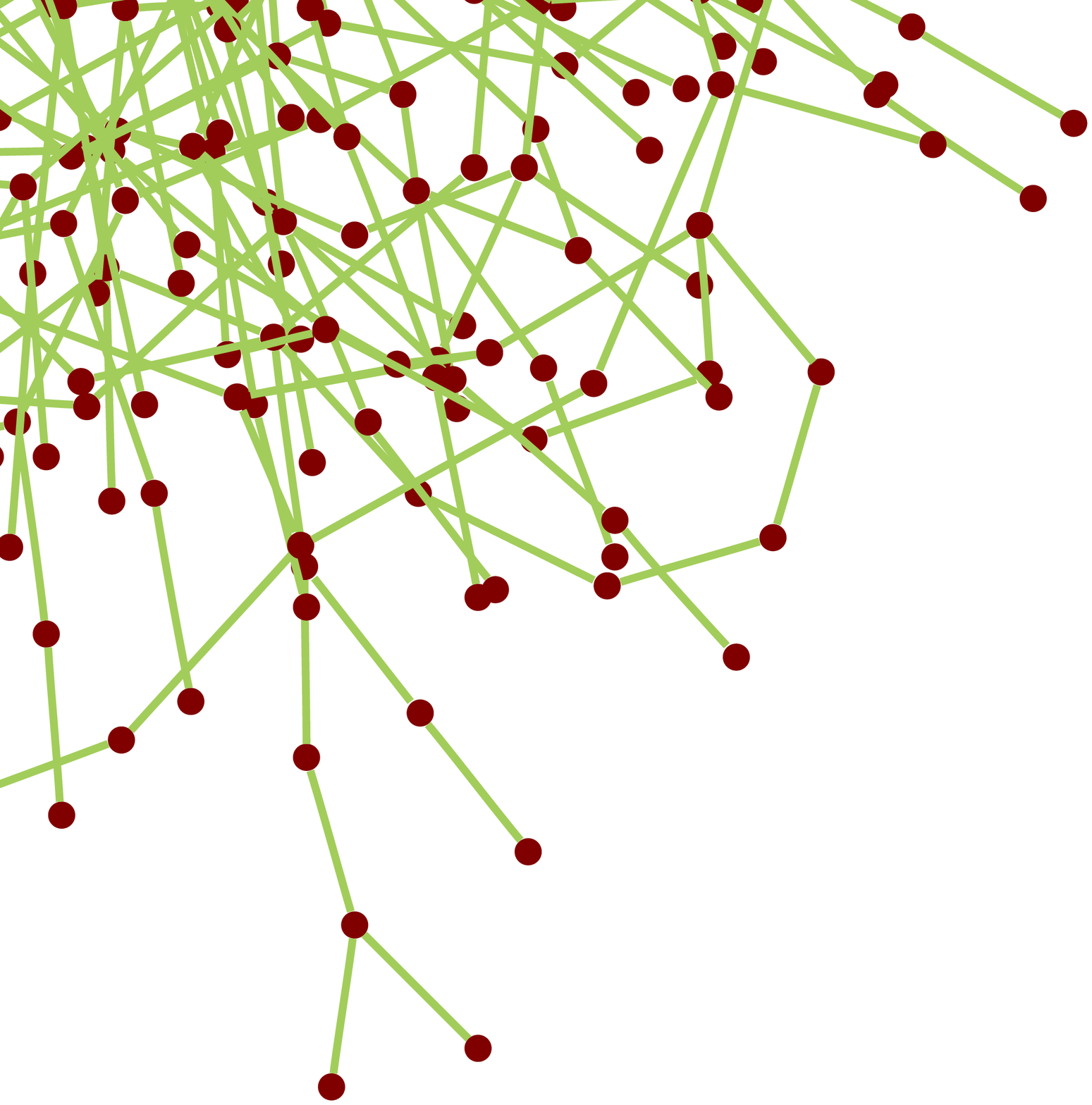}
        \label{fig:hot0k}}
        \hfill
        \subfigure[$1K$-graph]
        {\includegraphics[width=1.55in]{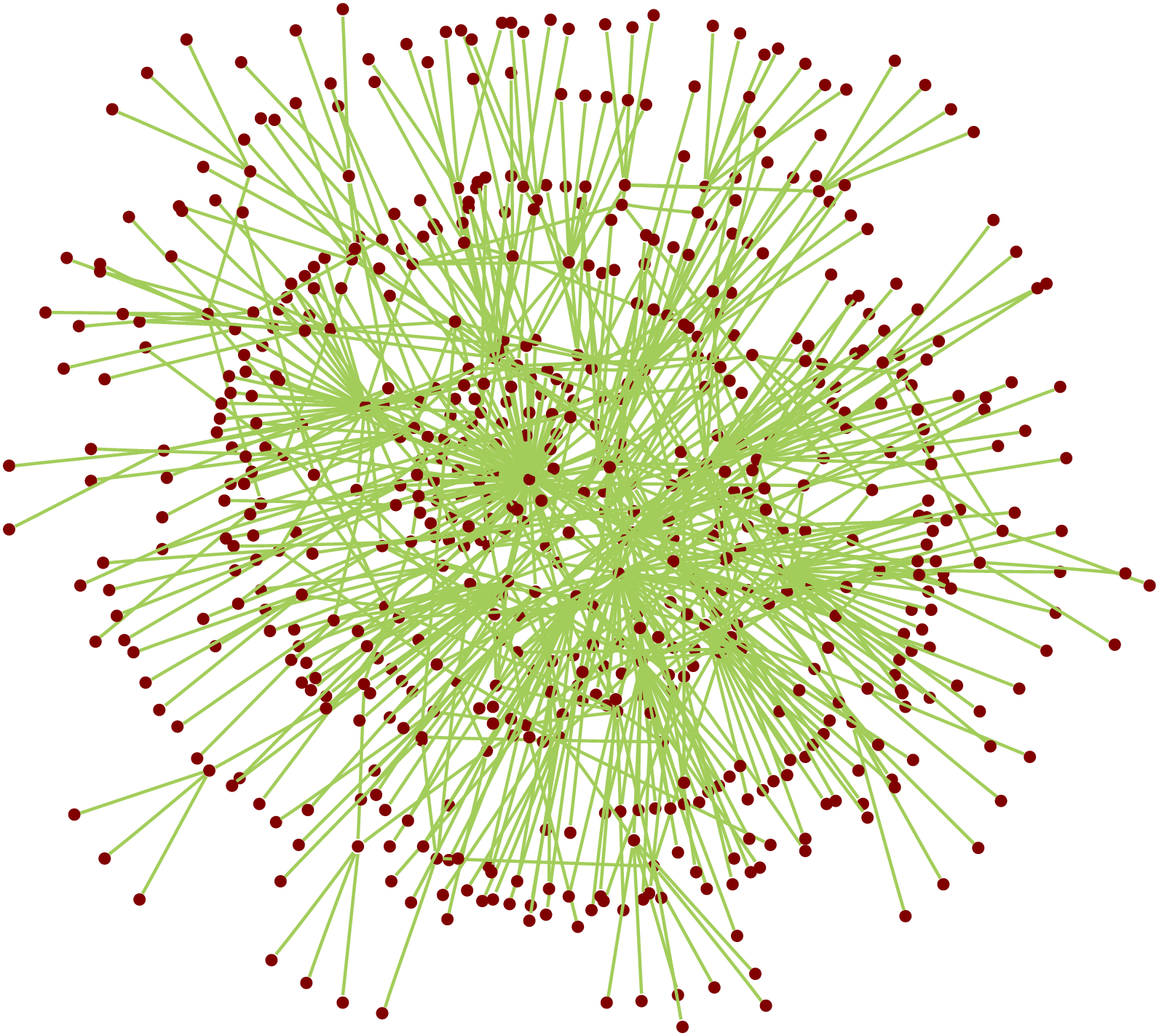}
        \label{fig:hot1k}}
        \hfill
        \subfigure[$2K$-graph]
        {\includegraphics[width=1.55in,angle=90]{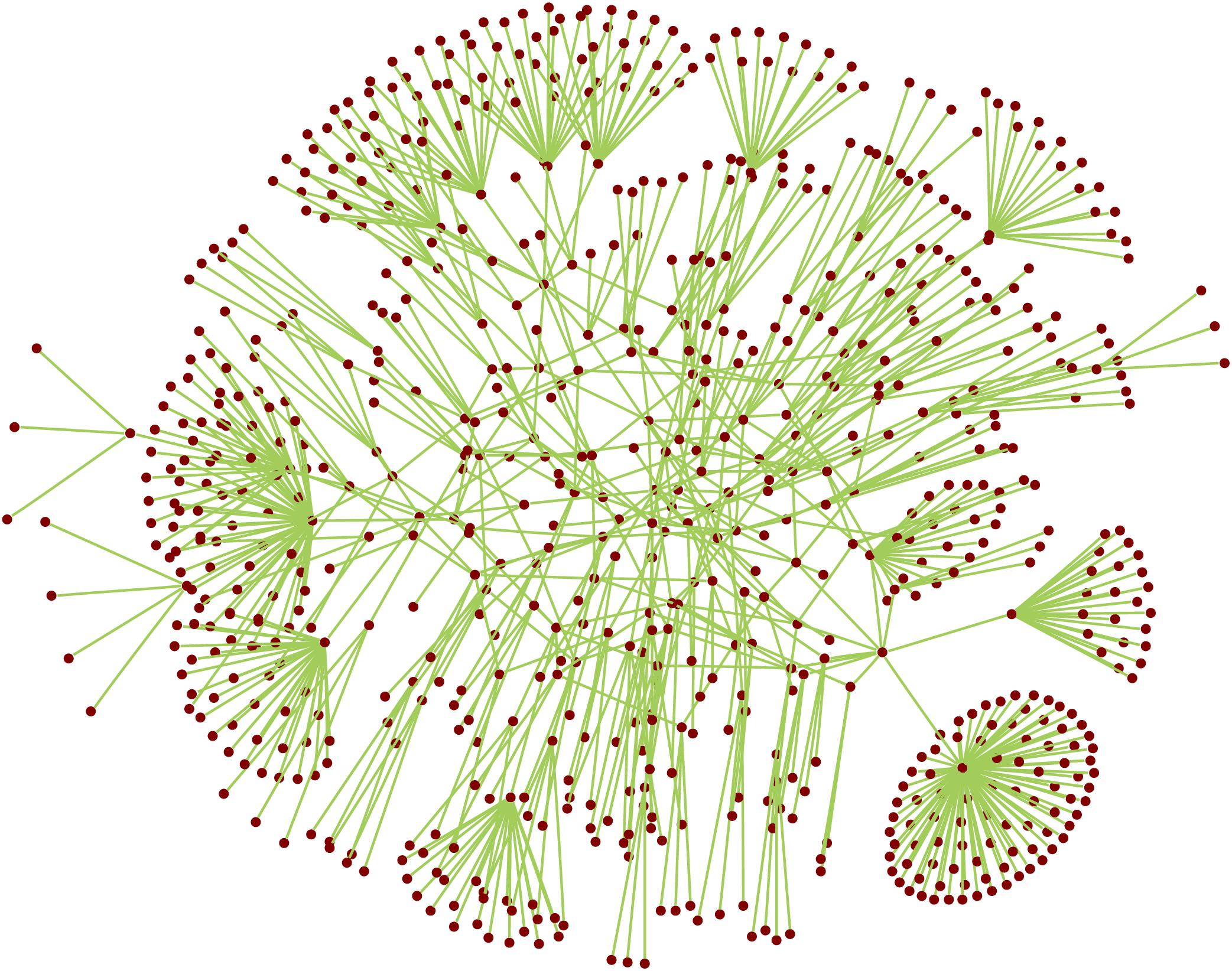}
        \label{fig:hot2k}}
    }
\vspace{-0.4in}
    \centerline{
        \subfigure[$3K$-graph]
        {\includegraphics[width=1.7in,angle=105]{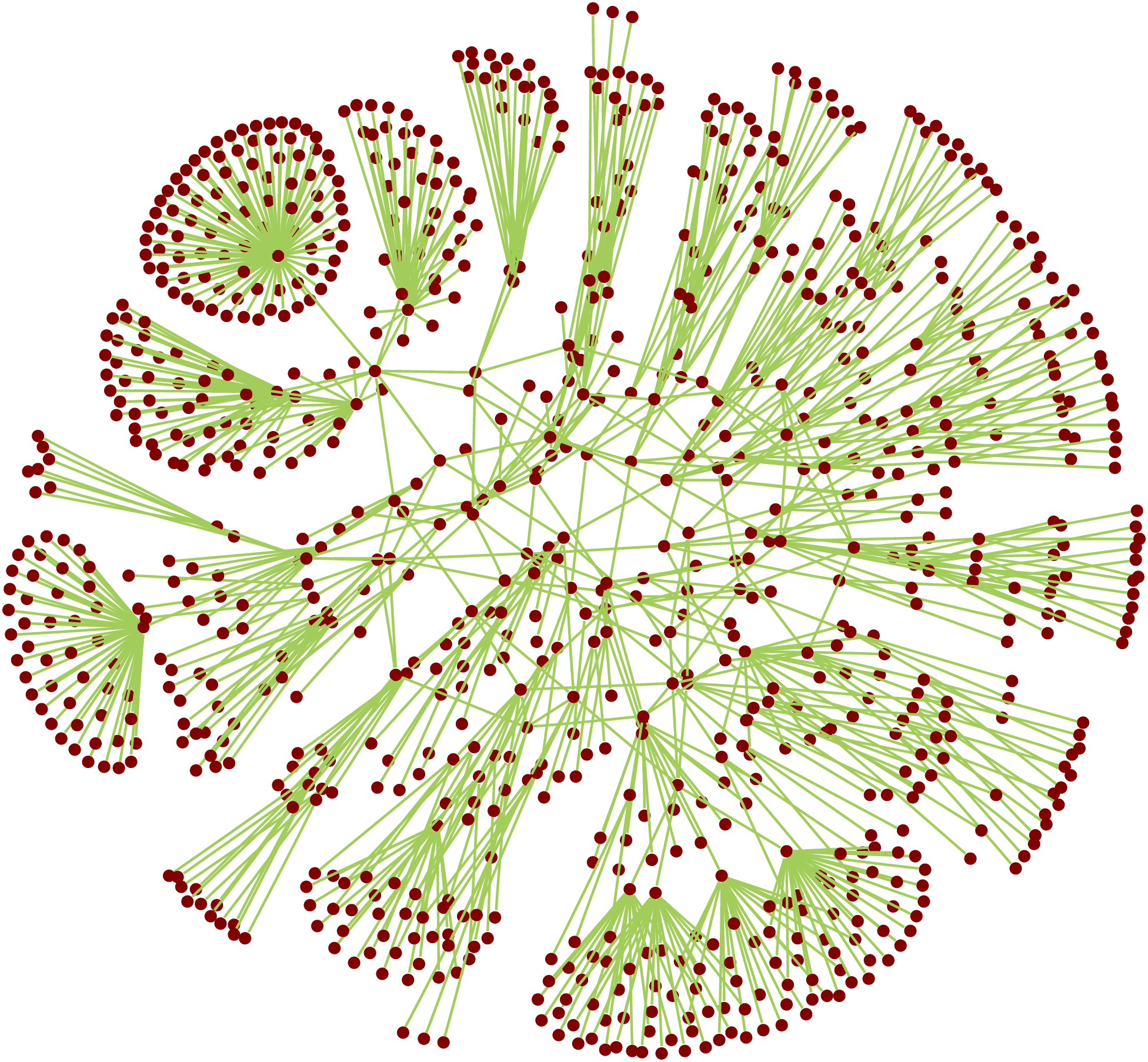}
        \label{fig:hot3k}}
        \hspace{0.65in}
        \subfigure[original HOT graph]
        {\includegraphics[width=1.65in,angle=-20]{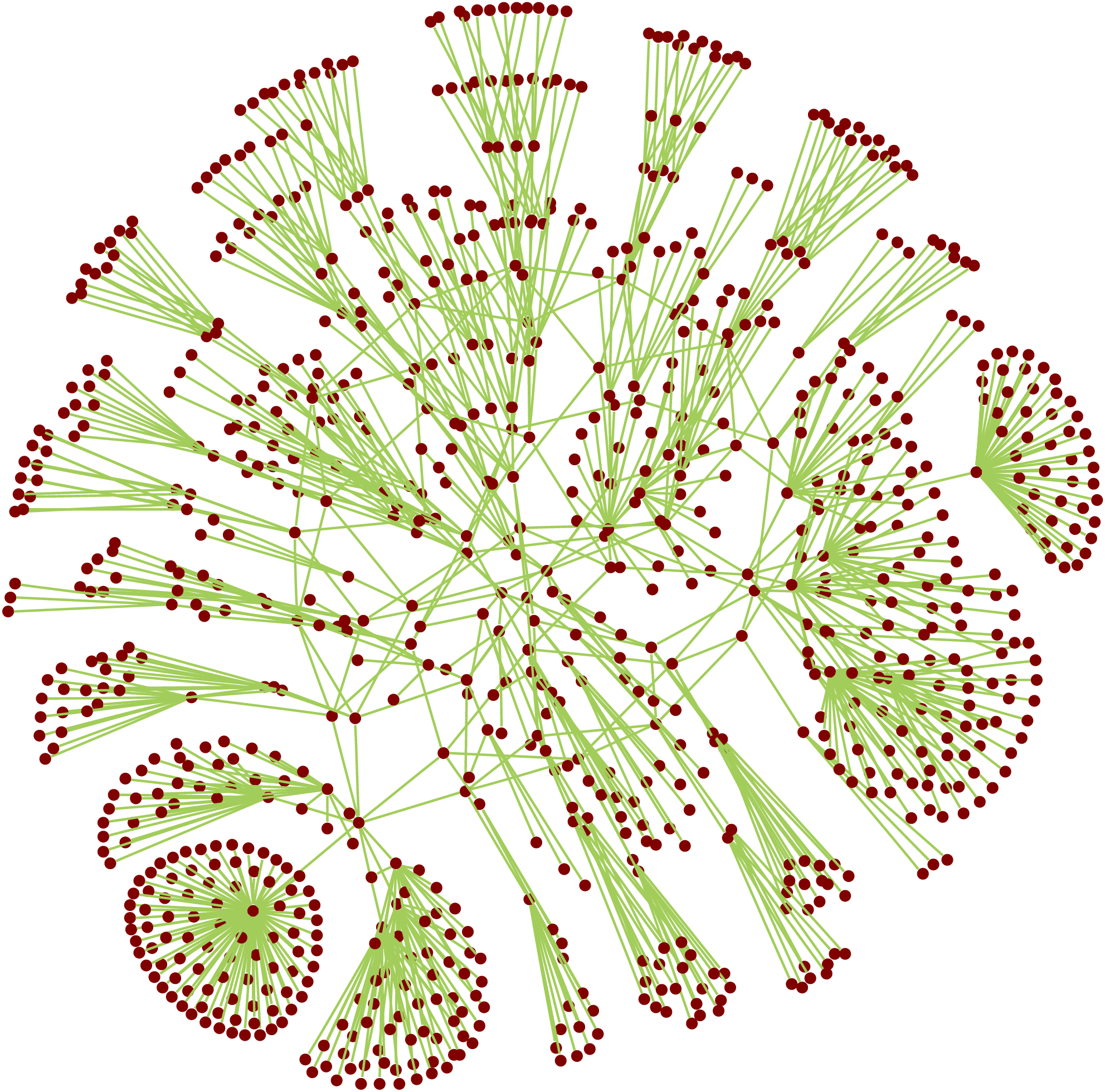}
        \label{fig:hot}}
        }
\vspace{-0.15in}
    \caption{\footnotesize {\bf Picturizations of $dK$-graphs and the original HOT graph
    illustrating the convergence of $dK$-series.}
}

    \label{fig:hot-cool-pic}
\end{figure*}

The $dK$-series definition satisfies the inclusion and convergence
requirements described above.  Indeed, the
inclusion requirement is satisfied because any graph of size~$d$
is a subgraph of some graph of size~\mbox{$d+1$}. Convergence follows from
the observation that in the limit of $d=n$, the set of $nK$-graphs
contains only one element:~$G$ itself.  As a consequence of the
convergence property, any topology metric we can define on~$G$ will
eventually be captured by $dK$-graphs with a sufficiently large~$d$.

Hereafter, our main concerns with the $dK$-series become: 1)~how well we
can satisfy our first requirement of constructibility and 2)~how fast
the $dK$-series converges toward the original graph. We address these two concerns
in Sections~\ref{sec:methodology} and~\ref{sec:results}.

The reason for the second concern is that the number of
probability distributions required to fully specify the $dK$-distribution
grows quickly with~$d$: see~\cite{sloane-A001349} for the number of
non-isomorphic simple connected graphs of size~$d$. Relative to the
existing work on topology generators typically limited to
$d=1$~\cite{AiChLu00,MeLaMaBy01,WiJa02},
we introduce and implement algorithms for graph construction
for ~$d=2$ and~$d=3$. We present these algorithms
in Section~\ref{sec:methodology} and then show in Section~\ref{sec:results}
that the $dK$-series converges quickly:
$2K$-graphs are sufficient for most practical purposes for the graphs
we consider, while
$3K$-graphs are essentially identical to observed and modeled Internet
topologies.

To motivate our ability to capture increasingly complex graph properties
by increasing~$d$, we present visualizations of $dK$-graphs generated using
the $dK$-randomizing approach we will discuss in Section~\ref{subsec:dk-rewiring}.
Figure~\ref{fig:hot-cool-pic} depicts random $0K$-, $1K$-, $2K$-
and $3K$-graphs matching the corresponding
distributions of the HOT graph, a representative
router-level topology from~\cite{LiAlWiDo04}. This topology is particularly interesting, because, to date reproducing router-level topologies using only
degree distributions has proven difficult~\cite{LiAlWiDo04}.
However, a visual inspection of our generated topologies shows good convergence
properties of the $dK$-series: while the $0K$-graph and $1K$-graph
have little resemblance with the HOT topology, the $2K$-graph is much closer than
the previous ones and the $3K$ graph is almost identical to the original.
Although the visual inspection is encouraging, we defer more careful
comparisons to Section~\ref{sec:results}.

\section{Constructing {\it\rm\large\lowercase{d}K}-graphs}
\label{sec:methodology}
There are several approaches for constructing $dK$-graphs for
\mbox{$d=0$} and \mbox{$d=1$}. We extended a number of these
algorithms to work for higher values of~$d$. In
Section~\ref{sec:dk-constructions}, we describe these approaches,
their practical utility, and our new algorithms for~\mbox{$d>1$}.
In Sections~\ref{sec:dk-random-graphs}
and~\ref{sec:dk-exploraitons}, we introduce {\em $dK$-random graphs\/}
and {\em $dK$-space explorations}. We use the latter to
determine the lowest values of~$d$ such that $dK$-graphs approximate a
given topology with the required degree of accuracy.

\subsection{$dK$-graph-constructing algorithms}
\label{sec:dk-constructions}

We classify existing approaches to constructing $0K$- and $1K$-graphs
into the following categories: {\em stochastic}, {\em pseudograph},
{\em matching}, and two types of {\em rewiring}: {\em randomizing\/}
and {\em targeting}.  We attempted to extend each of these techniques
to general $dK$-graph construction.  In this section, we qualitatively
discuss the relative merits of each of these approaches before
presenting a more quantitative comparison in Section~\ref{sec:results}.

\subsubsection{Stochastic}
\label{sec:dk-stochastic}

The simplest and most convenient for theoretical analysis is the
stochastic approach. For $0K$, reproducing an $n$-sized graph with a
given expected average degree~$\bar{k}$ involves connecting every pair
of $n$~nodes with probability~\mbox{$p_{0K}=\bar{k}/n$}.  This
construction forms the classical (Erd\H{o}s-R\'{e}nyi) random
graphs~$\mathcal{G}_{n,p}$~\cite{ErRe59}.  Recent efforts have
extended this stochastic approach to $1K$~\cite{ChLu02} and
$2K$~\cite{BoPa03,dorogovtsev03}.  In these cases, one
first labels all nodes~$i$ with their expected degrees~$q_i$ drawn
from the distribution~$P(k)$ and then connects pairs of nodes~$(i,j)$
with probabilities $p_{1K}(q_i,q_j) = q_iq_j/(n\bar{q})$ or
$p_{2K}(q_i,q_j) = (\bar{q}/n) P(q_i,q_j)/(P(q_i)P(q_j))$ reproducing
the expected values of~$1K$- or~$2K$-distributions,
respectively.

In theory, we could generalize this
approach for any~$d$ in two stages: 1)~{\em extraction}: given a
graph~$G$, calculate the frequencies of all (including disconnected)
$d$-sized subgraphs in $G$, and 2)~{\em construction}: prepare an
$n$-sized set of $q_i$-labeled nodes and connect their $d$-sized
subsets into different subgraphs with (conditional) probabilities based on the
calculated frequencies.  In practice, we find the stochastic
approach performs poorly even for $1K$ because of high statistical
variance. For example, many nodes with expected degree~$1$ wind
up with degree~$0$ after the construction phase, resulting in
many tiny connected components.

\subsubsection{Pseudograph}
\label{sec:dk-pseudograph}

The pseudograph (also known as {\em configuration}) approach is
probably the most popular and widely used class of
graph-generating algorithms.
In its original form~\cite{AiChLu00,MolRee95}, it applies only to the
$1K$ case.  Relative to the stochastic approach, it reproduces a given
degree distribution exactly, but does not necessarily construct
simple graphs.  That is, it may construct graphs with both ends of
an edge connected to the same node (self-loops) and with multiple
edges between the same pair of nodes (loops).

It operates as follows. Given the number of nodes, $n(k)$, of degree $k$,
$n=\sum_{k=1}^{k_{\max}}n(k)$, first prepare $n(k)$~nodes with
$k$~stubs attached to each node, $k=1,\ldots,k_{\max}$, and then
randomly choose pairs of stubs and connect them to form edges. To
obtain a simple connected graph, remove all loops and extract the
largest connected component.

We extended this algorithm to $2K$ as follows. Given the
number~$m(k_1,k_2)$ of edges between $k_1$- and $k_2$-degree
nodes, $m=\sum_{k_1,k_2=1}^{k_{\max}}m(k_1,k_2)$, we first prepare
lists of~$m(k_1,k_2)$ disconnected edges and label the both ends of
each edge by~$k_1$ and~$k_2$, $k_1,k_2=1,\ldots,k_{\max}$.
Next, for each~$k$, $k=1,\ldots,k_{\max}$, we create a list of all
edge-ends labeled with~$k$. From this list, we randomly select groups
of $k$~edge-ends to form the $k$-degree nodes in the final graph.

The pseudograph algorithm works well for~\mbox{$d=2$}.
Unfortunately, we could not easily generalize it for~\mbox{$d>2$}
because starting at~\mbox{$d=3$},
$d$-sized subgraphs overlap over edges. Such overlapping introduces a series
of topological constraints and non-local dependencies among different
subgraphs, and we could not find a simple technique to
preserve these combinatorial constraints during the construction phase.

\subsubsection{Matching}

The matching approach differs from the pseudograph approach in
avoiding loops during the construction phase. In the $1K$ case,
the algorithm works exactly as its pseudograph counterpart but skips
pairs of stubs that form loops if connected.
We extend the matching approach to~$2K$ in a manner similar to our
$2K$~pseudograph approach.

Unfortunately, loop avoidance suffers from various forms of
deadlock for both $1K$ and $2K$. In both cases, the algorithms can end up in
incomplete configurations when not all edges are formed, and the
graph cannot be completed because there are no suitable stub pairs remaining
that can be connected without forming loops. We devised several
techniques to deal with these problems.
With these additional techniques, we obtained good results for $2K$ graphs.
As in the pseudograph case however, we could not generalize matching
for~\mbox{$d>2$} for essentially the same reasons
related to subgraphs' overlapping and non-locality.

\subsubsection{Rewiring}
\label{subsec:dk-rewiring}
\begin{figure}
    \centerline{\includegraphics[width=2.5in]{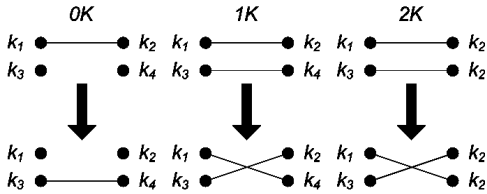}}
    \caption{\footnotesize \bf $dK$-preserving rewiring for~\mbox{$d=0,1,2$}.}
    \label{fig:dk-rewiring}
\end{figure}

The rewiring approaches are generalizable to any~$d$ and
work well in practice.  They involve $dK$-preserving rewiring
as illustrated in Figure~\ref{fig:dk-rewiring}.
The main idea is to rewire {\em random\/} (pairs of) edges
preserving an existing form of the $dK$-distribution. For
\mbox{$d=0$}, we rewire a random edge to a random pair of nodes,
thus preserving~$\bar{k}$.  For~\mbox{$d=1$}, we rewire two random
edges that do not alter~$P(k)$, as shown in Figure~\ref{fig:dk-rewiring}. If, in
addition, there are at least two nodes of equal degrees adjacent to the
different edges in the edge pair, then the same rewiring
leaves~$P(k,k')$ intact.  Due to the inclusion property of the
$dK$-series, \mbox{$(d+1)K$}-rewirings form a subset
of~\mbox{$dK$}-rewirings for~\mbox{$d>0$}. For example, to
preserve~$3K$, we permit a $2K$-rewiring only if it also preserves the
wedge and triangle distributions.

The {\em $dK$-randomizing rewiring algorithm\/} amounts to performing
$dK$-preserving rewirings a sufficient number of times for some
$dK$-graph.
A ``sufficient number'' means enough rewirings for this process to lead to
graphs that do not change their properties even if we subject them to
additional rewirings.
In other words, this rewiring process {\em converges}
after some number of steps, producing
random graphs having property~$\mathcal{P}_d$.
Even for~\mbox{$d=1$}, there are no known rigorous results
regarding how quickly this process converges, but~\cite{GkaMiZe03a}
shows that this process is an irreducible, symmetric and aperiodic
Markov chain and demonstrates experimentally that it takes $O(m)$~steps
to converge.

In our
experiments in Section~\ref{sec:results}, we employ the following
strategy applicable for any~$d$. We first calculate the number of
possible initial $dK$-preserving rewirings. By ``initial rewirings'' we
mean rewirings we can perform on a given graph~$G$, to differentiate
them from rewirings we can apply to graphs obtained from~$G$ after its
first (and subsequent) rewirings.  We then subtract the number of
rewirings that leave the graph isomorphic. For example, rewiring of
any two \mbox{$(1,k)$}- and \mbox{$(1,k')$}-edges is a $dK$-preserving
rewiring, for any~$d$, and more strongly, the graph before rewiring is
isomorphic to the graph after rewiring.  We multiply this difference by~$10$,
and perform that number of random rewirings.  At
the end of our rewiring procedure, we explicitly verify that
randomization is indeed complete and the process has converged by further increasing the number of rewirings and checking that
all graph characteristics remain unchanged.

One obvious problem with $dK$-randomization
is that it requires an original graph~$G$ as input to construct its $dK$-random versions.
It cannot start with a description of the $dK$-distribution
to generate random $dK$-graphs as is possible with the
other construction approaches discussed above.

To address this limitation, we consider the inverse process of {\em
$dK$-targeting $d'K$-preserving rewiring}, also known as {\em
Metropolis dynamics\/}~\cite{Metro53}.  It incorporates the
following modification to $d'K$-preserving rewiring: every rewiring
step is accepted only if it moves the graph ``closer''
to~$\mathcal{P}_{d}$.
In practice, we can then employ targeting rewiring to construct
$dK$-graphs with high values of~$d$ by beginning with any $d'K$-graph
where~\mbox{$d'<d$}. Recall that we can always compute~$\mathcal{P}_{d'}$
from~$\mathcal{P}_{d}$  due to the inclusion
property of the $dK$-series. For instance, we can start with a graph
having a given degree distribution
(\mbox{$d'=1$})~\cite{ViLa05-cocoon}, and then move it toward
a~$dK$-graph via $dK$-targeting $1K$-preserving rewiring.

The definition of ``closer'' above requires further explanation.
We require a set of distance metrics that
quantitatively differentiate two graphs based on the values
of their $dK$-distributions.
In our experiments, we use the sum of squares of differences
between the existing and target numbers of subgraphs of a given
type. For example, in the \mbox{$d=2$}~case, we measure the distance
between the target graph's JDD and the JDD of the current graph being rewired by $\mathcal{D}_2=\sum_{k_1,k_2}
  \left[m_{\mathrm{current}}(k_1,k_2)-m_{\mathrm{target}}(k_1,k_2)\right]^2$,
and at each rewiring step, we accept the rewiring only if it decreases
this distance.  Note that $\mathcal{D}_2$ is non-negative and equals
zero only when reaching the target JDD. For \mbox{$d=3$}, this distance~$\mathcal{D}_3$
is a sum of squares of differences between the current and target numbers of
wedges and triangles, and we can generalize it to~$\mathcal{D}_d$ for any~$d$.

A potential problem with $dK$-targeting
$d'K$-preserving rewiring is that it can be nonergodic, meaning that
there might be no chain of $d'K$-preserving $\mathcal{D}_d$-decreasing rewirings leading from the
initial $d'K$-graph to the target $dK$-graph. In other
words, we cannot be sure beforehand that any two $d'K$-graphs are connected by
a sequence of $d'K$-preserving and $\mathcal{D}_d$-decreasing rewirings.

To address this problem we
note that the $d'K$-randomizing and $dK$-targeting $d'K$-preserving
rewirings are actually two extremes of an entire family of rewiring
processes. Indeed, let \mbox{$\Delta \mathcal{D}_d =
  \mathcal{D}_{d,\mathrm{after}}-\mathcal{D}_{d,\mathrm{before}}$} be
the difference of distance to the target $dK$-distribution computed
before and after a $d'K$-preserving rewiring step.  As with the usual
$dK$-targeting rewiring, we accept a rewiring step if~\mbox{$\Delta
\mathcal{D}_d < 0$}, but even if~\mbox{$\Delta \mathcal{D}_d \geq 0$},
we also accept this step with probability $e^{-\Delta\mathcal{D}_d/T}$,
where~$T>0$ is some parameter that we call {\em
temperature\/} because of the similarity of the process to
simulated annealing.

In the \mbox{$T \to 0$}~limit, this probability goes to~$0$, and we
have the standard $dK$-targeting $d'K$-preserving rewiring
process. When~\mbox{$T \to \infty$}, the probability approaches~$1$,
yielding the standard $d'K$-randomizing rewiring process. To verify
ergodicity, we can start with a high temperature and then gradually
cool the system while monitoring any metric known to have different
values in $dK$- and $d'K$-graphs. If this metric's value forms a
continuous function of the temperature, then our rewiring process is
ergodic. Maslov {\it et al.}~performed these experiments
in~\cite{MaSneZa04} and demonstrated ergodicity in the case with
\mbox{$d'=1$} and \mbox{$d=2$}. In our experiments in
Section~\ref{sec:results} where~$d$ and~$d'$ are below~$4$,
we always obtain a good match for all target graph metrics.
Thus, we perform rewiring at zero temperature without
further considering ergodicity. If however in some future experiments
one detects the lack of a smooth convergence of rewiring procedures,
then one should first verify ergodicity using the methodology described above.\\

For all the algorithms discussed in this section, we do not check for graph
connectedness at
each step of the algorithm
since: 1)~it is an expensive operation
and 2)~all resulting graphs always have giant connected components~(GCCs)
with characteristics similar to the whole disconnected graphs.

\subsection{$dK$-random graphs}
\label{sec:dk-random-graphs}

\begin{table*}[tb]
\begin{centering}
\caption{The summary of $dK$-series.
\label{tab:dk}
}
\begin{tabular}{|p{.15in}|p{.45in}|p{.73in}|p{1.02in}|p{1.65in}|p{2.05in}|}\hline
Tag $dK$ & Property symbol & $dK$-distribution & $\mathcal{P}_d$ defines $\mathcal{P}_{d-1}$ &
Edge existence probability in stochastic constructions
& Maximum entropy value of $(d+1)K$-distribution in $dK$-random graphs\\
\hline $0K$ & $\mathcal{P}_0$  & $\bar{k}$ & & $p_{0K}=\bar{k}/n$
& $P_{0K}(k) = e^{-\bar{k}}\bar{k}^k/k!$\\
\hline $1K$ & $\mathcal{P}_1$ & $P(k)$ & $\bar{k}=\sum kP(k)$ & $p_{1K}(q_1,q_2)
= q_1q_2/(n\bar{q})$
& $P_{1K}(k_1,k_2)=k_1P(k_1)k_2P(k_2)/\bar{k}^2$\\
\hline $2K$ & $\mathcal{P}_2$ & $P(k_1,k_2)$ & $P(k) =
(\bar{k}/k)\sum_{k'}P(k,k')$ & $p_{2K}(q_1,q_2) = (\bar{q}/n)
P(q_1,q_2)/(P(q_1)P(q_2))$ & See~\cite{dorogovtsev04} for clustering in $2K$-random graphs \\
\hline $3K$ & $\mathcal{P}_3$ & $P_\wedge(k_1,k_2,k_3)$ $P_\triangle(k_1,k_2,k_3)$ &
\multicolumn{3}{l|}{\parbox[t]{4.72in}{
By counting edges, we get
$P(k_1,k_2) \sim \sum_k \left\{ P_\wedge(k,k_1,k_2) + P_\triangle(k,k_1,k_2) \right\} / (k_1-1)
\sim \sum_k \left\{ P_\wedge(k_1,k_2,k) + P_\triangle(k_1,k_2,k) \right\} / (k_2-1)$,
where we omit normalization coefficients.}}\\
\hline
\ldots&\ldots&\ldots&\ldots&\ldots&\ldots\\
\hline
$nK$ & $\mathcal{P}_n$ & $G$ & & &\\
\hline
\end{tabular}
\end{centering}
\end{table*}

No $dK$-graph-generating algorithm can quickly construct the set of
{\em all\/} $dK$-graphs because: 1)~such sets are too large,
especially for small~$d$; and, less obviously, 2)~all algorithms try
to produce graphs having property~$\mathcal{P}_d$ while remaining {\em
  unbiased\/} (random) with respect to all other properties.  One can
check directly that the last characteristic applies to all the
algorithms we have discussed above.

As a consequence, the $dK$-graph construction algorithms
result in non-uniform sampling of graphs with different values of properties
that are not fully defined by~$\mathcal{P}_d$. More specifically, two generated $dK$-graphs
having different forms of a $d'K$-distribution with~\mbox{$d'>d$}
can appear as the output of these
algorithms with drastically different probabilities. Some $dK$-graphs
have such a small probability of being constructed
that we can safely assume they never arise.

For example, consider the simplest $0K$ stochastic
construction, i.e., the classical random
graphs~$\mathcal{G}_{n,p}$. Using a probabilistic argument, one can
show that the naturally-occurring $1K$-distribution~(degree distribution)
in these gra\-phs has a specific form: binomial, which is
closely approximated by the Poisson distribution: \mbox{$P_{0K}(k) =
  e^{-\bar{k}}\bar{k}^k/k!$}~\cite{DorMen-book03}.
The $0K$ stochastic algorithm may produce a graph with a different
$1K$-distribution, e.g., the
power-law~\mbox{$P(k) \sim k^{-\gamma}$}, but the probability of such
an outcome is extremely low.
Indeed, suppose \mbox{$n \sim 10^4$},
\mbox{$\bar{k} \sim 5$}, and \mbox{$\gamma \sim 2.1$}, so
that the characteristic maximum degree is \mbox{$k_{\max} \sim 2000$}
(we chose these values to reflect measured values for Internet AS
topologies).
In this case, the probability that a
$\mathcal{G}_{n,p}$-graph contains at least one node with degree equal
to~$k_{\max}$ is dominated by \mbox{$1/2000! \sim 10^{-6600}$}, and
the probability that the remaining degrees simultaneously match those
required for a power law is much lower.

It is thus natural to introduce a set of graphs that correspond
to the graphs most likely to be generated by $dK$-graph constructing
algorithms.
We call such graphs the {\em $dK$-random graphs}.
These graphs have property~$\mathcal{P}_d$ but
are unbiased with respect to any other more constraining property.
In this sense, the
$dK$-random graphs are the {\em maximally random\/} or {\em
  maximum-entropy\/} $dK$-graphs. Our term {\em maximum entropy\/}
here has the following justification.  As we have just seen,
$0K$-random graphs have the maximum-entropy value of
the $1K$-distribution since their node degree distribution is the
distribution with the maximum entropy among all the distributions with
a fixed average.\footnote{The entropy of a discrete
  distribution~$P(x)$ is $H[P(x)]=-\sum_x P(x) \log P(x)$. If the sample
  space is also finite, then among all the distributions with a fixed average,
  the binomial distribution maximizes entropy~\cite{harrem01}.}
The $1K$-random graphs have the
maximum-entropy value of the $2K$-distribution since their joint
degree distribution,
\mbox{$P_{1K}(k_1,k_2)=\tilde{P}(k_1)\tilde{P}(k_2)$}, where
\mbox{$\tilde{P}(k)=kP(k)/\bar{k}$}~\cite{DorMen-book03}, is the distribution with the
maximum joint entropy (minimum mutual information)\footnote{The mutual
  information of a joint distribution~$P(x,y)$ is
  $I[P(x,y)]=H[P(x)]+H[P(y)]-H[P(x,y)]$, where $P(x)$ and $P(y)$ are
  the marginal distributions.} among all the joint distributions with
fixed marginal distributions.\footnote{In reality, the last statement
  generally applies only to the class of all (not necessarily connected)
  pseudographs.  Narrowing the class of graphs to simple connected
  graphs introduces topological constraints affecting the
  maximum-entropy form of the $2K$-distribution.}

The main point we extract from these observations is that in trying to
construct $dK$-graphs, we generally obtain graphs from subsets
of the $dK$-space.
We call these subsets $dK$-random graphs and
schematically depict them as centers of the $dK$-circles in
Figure~\ref{fig:dk}. These graphs do have property~$\mathcal{P}_d$ and,
consequently, properties~$\mathcal{P}_i$ with \mbox{$i<d$},
but they might not ever display property~$\mathcal{P}_j$ with \mbox{$j>d$}
since their $jK$-distributions have specific, maximum-entropy values
that may be different from the $jK$-distribution values in the original graph.

\subsection{$dK$-space explorations}
\label{sec:dk-exploraitons}

Often we wish to analyze the topological constraints
a given graph~$G$ appears to obey.
In other cases, we are interested in
exploring the structural diversity among $dK$-graphs.
If we are attempting to determine the minimum~$d$ such that
all $dK$-graphs are similar to~$G$,
we can start with a small value of~$d$, generate $dK$-graphs, and
measure their ``distance'' from~$G$.
If the distance is too great, we can increase~$d$ and repeat
the process.
On the other hand, to explore structural diversity
among all $dK$-graphs, we must generate $dK$-graphs that are not random.
These non-random $dK$-graphs are still constrained by~$\mathcal{P}_d$ but
have extremely low probabilities of being generated by unperturbed
$dK$-graph constructing algorithms.

We cannot construct all $dK$-graphs, so we need to use
heuristics to generate some $dK$-graphs and adjust them
according to a distance metric that draws us closer to
the types of $dK$-graphs we seek.  One such
heuristic is based on the inclusion feature of the
$dK$-series. Because all $dK$-graphs have the same values
of $dK$- but not of $(d+1)K$-distributions, we look for simple metrics
fully defined by~$\mathcal{P}_{d+1}$ but not by~$\mathcal{P}_d$.
While identifying such metrics can be
challenging for high~$d$'s, we can always retreat to the following two
simple extreme metrics:
\begin{itemize}
\item the correlation of degrees of nodes located at distance~\mbox{$d$};
\item the concentration of $d$-simplices (cliques of size~\mbox{$d+1$}).
\end{itemize}
These metrics are ``extreme'' in the sense that they correspond
to the \mbox{$(d+1)$}-sized subgraphs with, respectively,
the maximum~($d$) and minimum~($1$) possible diameter.
We can then construct $dK$-graphs with extreme values,
e.g., the smallest or largest possible, for these (extreme) metrics. The
$dK$-random graphs have the values of these metrics lying somewhere in between the extremes.

If the goal is to find the smallest~$d$ that results in sufficiently
constraining graphs, we can compute the difference between the extreme values of
these metrics, as well as of other metrics we might consider.  If this
difference is too large, then the selected value of~$d$ is not
constraining enough and we need to increase it.
If the goal is to visit exotic locations in a large space of $dK$-graphs, then
such $dK$-space exploration may be used to move beyond the relatively
small circle of $dK$-random graphs.

To illustrate this approach in practice, we consider $1K$-
and $2K$-space explorations. For $1K$, the simplest metric defined
by~$\mathcal{P}_2$ is any scalar summary statistics of the $2K$-distribution,
such as likelihood~$S$ (cf.~Section~\ref{sec:metrics}).  To construct graphs
with the maximum value of~$S$, we can run a form of targeting
$1K$-preserving rewiring that accepts each rewiring step only if it
increases~$S$. We can perform the opposite to minimize~$S$. This type
of experiment was at the core of recent work that led the authors
of~\cite{LiAlWiDo04} to conclude that \mbox{$d=1$} was not
constraining enough for the topology they considered.

To perform $2K$-space explorations, we need to find simple scalar
metrics defined by~$\mathcal{P}_3$.  Since the $3K$-distribution is
actually two distributions, $P_\wedge(k_1,k_2,k_3)$ and
$P_\triangle(k_1,k_2,k_3)$, we should have two independent scalar
metrics. The {\em second-order likelihood}~$S_2$ is one such metric
for~$P_\wedge(k_1,k_2,k_3)$. We define it as the sum of the products
of degrees of nodes located at the ends of wedges,
$S_2 \sim \sum_{k_1,k_2,k_3} k_1k_3P_\wedge(k_1,k_2,k_3)$.
Any graphs with the same~$P_\wedge(k_1,k_2,k_3)$ have the same~$S_2$.
For the $P_\triangle(k_1,k_2,k_3)$ component, average clustering
$\bar{C} \sim \sum_{k_1,k_2,k_3}k_1P_\triangle(k_1,k_2,k_3)$ is an
appropriate candidate. We note that these two metrics are also
the two extreme metrics in the sense defined above: $S_2$~measures the
properly normalized correlation of degrees of nodes located at distance~$2$,
while $\bar{C}$~describes the concentration of $2$-simplices (triangles).
The $2K$-explorations amount then to performing
the following two types of targeting~$2K$-preserv\-ing rewiring: accept a $2K$-rewiring
step only if it maximizes or minimizes: 1)~$S_2$, or 2)~$\bar{C}$.

\section{Evaluation}
\label{sec:results}
In this section, we conduct a number of experiments to demonstrate the
ability of the $dK$-series to capture important graph properties.
We implemented all the $dK$-graph-constructing algorithms from
Section~\ref{sec:dk-constructions}, applied them to both measured and
modeled Internet topologies, and calculated all
the topology metrics from Section~\ref{sec:metrics}
on the resulting graphs.

We experimented with three measured AS-level topologies, extracted
from CAIDA's {\em skitter\/} traceroute~\cite{as-adjacencies}, RouteViews'
{\em BGP\/}~\cite{routeviews}, and RIPE's {\em WHOIS\/}~\cite{irr}
data for the month of March 2004, as well as with a synthetic
router-level topology---the HOT graph from~\cite{LiAlWiDo04}.
The qualitative results of our experiments are similar for
the skitter and BGP topologies, while the WHOIS topology
lies somewhere in-between the skitter/BGP and HOT topologies.
In the case of skitter comprising of 9204 nodes and 28959 edges, we will see that its
degree distribution places significant constraints upon the graph
generation process.
Thus, even $1K$-random graphs approximate the skitter
topology
reasonably well. The HOT topology with 939 nodes and 988 edges is at the
opposite extreme. It is the least constrained; $1K$-random graphs approximate
it poorly, and $dK$-series' convergence is slowest.
We thus report results only for these two extreme cases, skitter and HOT.

Our results represent averages over 100 graphs generated with a different random seed
in each case, using the notation in Table~\ref{table:notations}.

\begin{table}
\caption{\footnotesize{\bf Scalar graph metrics notations.} }
\begin{tabular}{|l|l|}
\hline
Metric & Notation \\
\hline
Average degree & $\bar{k}$ \\
Assortativity coefficient & $r$ \\
Average clustering & $\bar{C}$ \\
Average distance & $\bar{d}$ \\
Standard deviation of distance distribution & $\sigma_{d}$ \\
Second-order likelihood & $S_2$ \\
Smallest eigenvalue of the Laplacian & $\lambda_{1}$ \\
Largest eigenvalue of the Laplacian & $\lambda_{n-1}$ \\
\hline
\end{tabular}
\label{table:notations}
\end{table}

\subsection{Algorithmic Comparison}
\label{subsec:algo-comp}

\begin{figure*}[tbh]
    \centerline{
    \subfigure[Clustering in skitter for different $2K$~algorithms]
        {\includegraphics[width=2.1in]{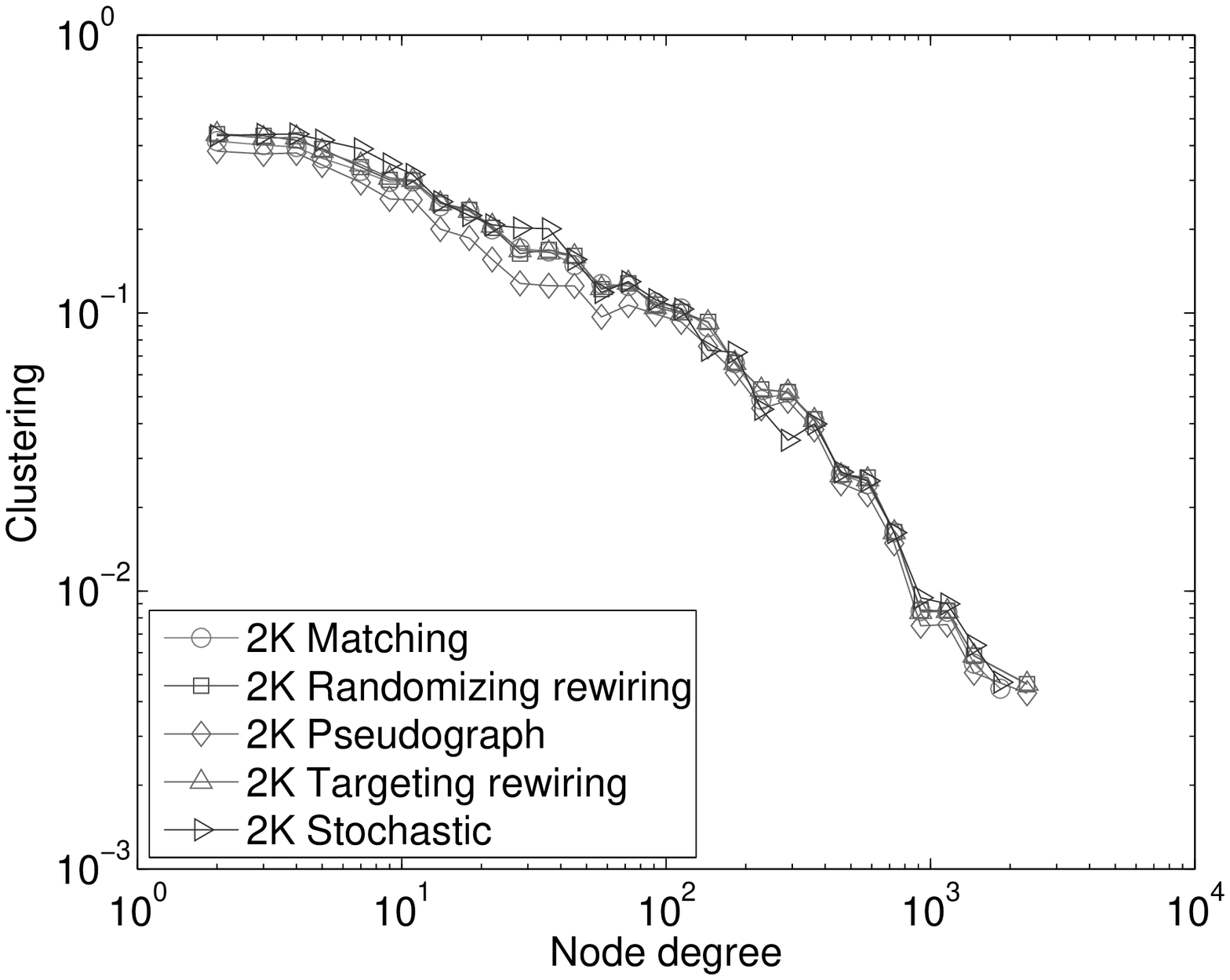}
        \label{fig:sk-all-clus} }
        \hfill
        \hfill
        \subfigure[Distance distribution in HOT for different $2K$~algorithms]
        {\includegraphics[width=2.1in]{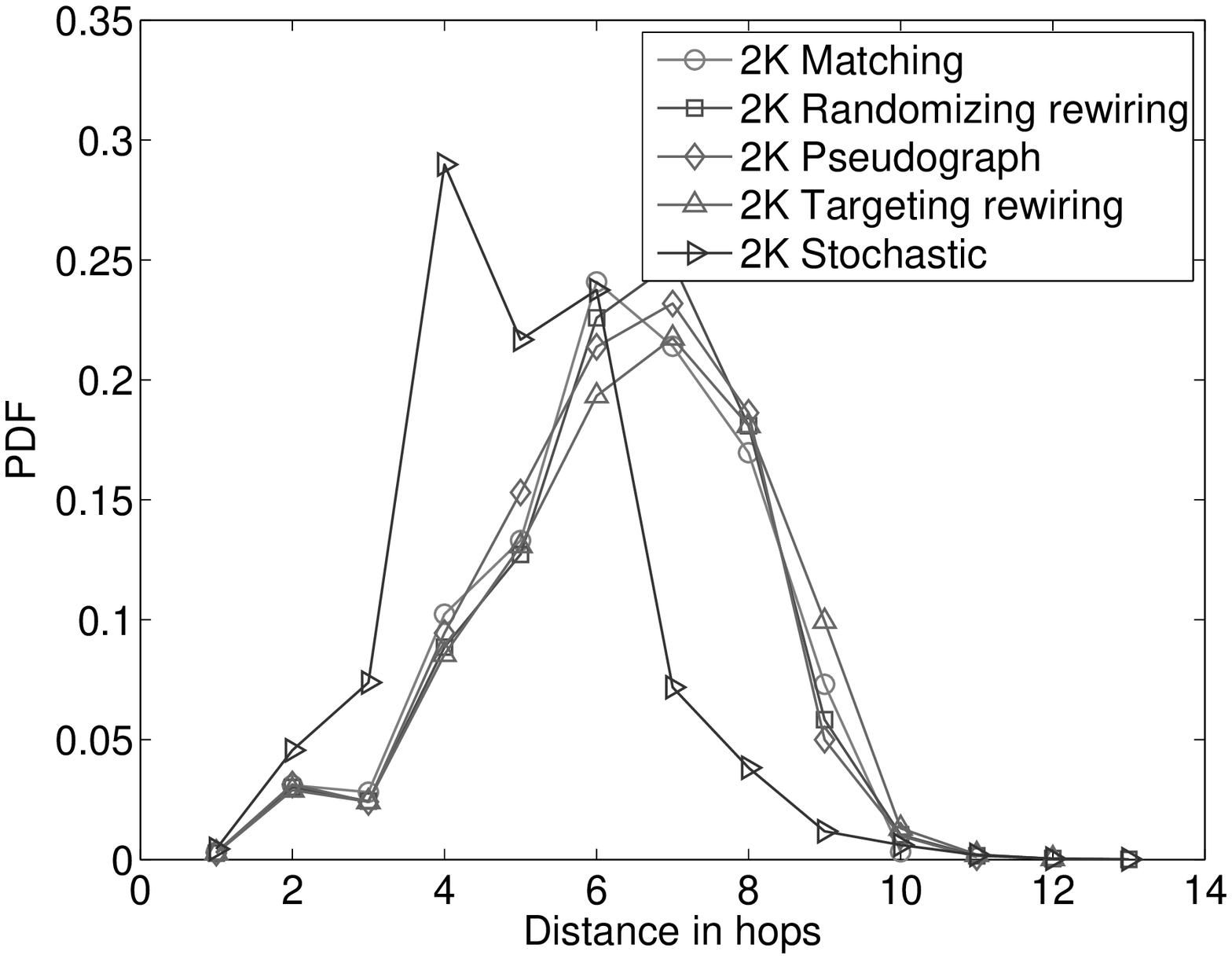}
        \label{fig:hot-all-dist} }
        \hfill
        \subfigure[Distance distribution in HOT for different $3K$~algorithms]
        {\includegraphics[width=2.1in]{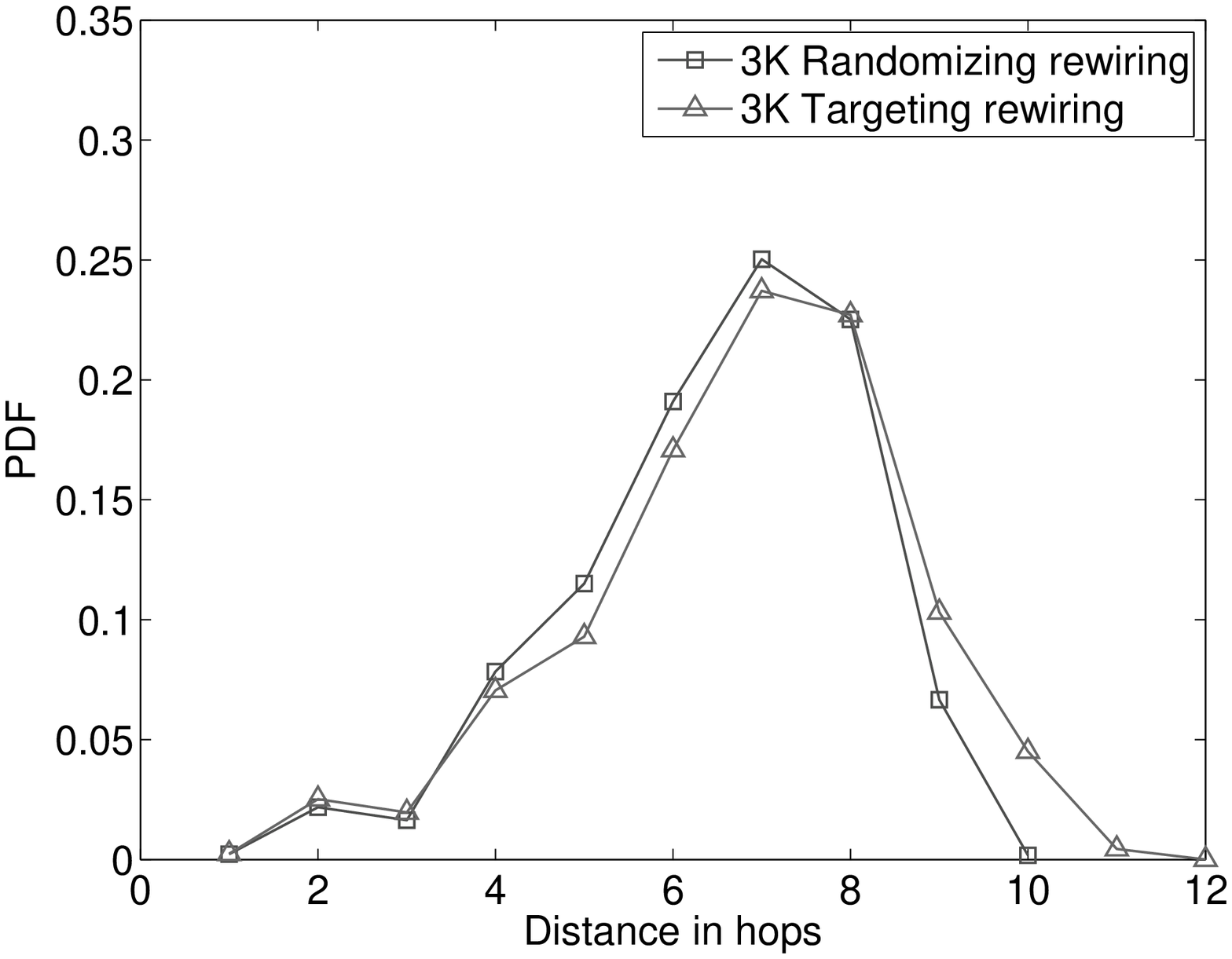}
        \label{fig:3k-dist-hot} }
    }
    \caption{\footnotesize {\bf Comparison of $2K$- and $3K$-graph-constructing algorithms.} }
    \label{fig:comp-all-algos}
\end{figure*}

\begin{table}
\begin{center}
\caption{\footnotesize{\bf Scalar metrics for $2K$-random HOT graphs generated using different techniques.}}
\begin{tabular}{|l|l|l|l|l|l|l|}
  \hline
  Met-          & Stoch- & Pseu-   & Match- & $2K$- & $2K$- & Orig. \\
  ric           & astic  & dogr.   & ing    & rand. & targ. & HOT   \\
  \hline
  $\bar{k}$     & 2.87   & 2.19    & 2.22   & 2.18  & 2.18  & 2.10  \\
  $r$           & -0.22  & -0.24   & -0.21  & -0.23 & -0.24 & -0.22 \\
  $\bar{d}$     & 4.99   & 6.25    & 6.22   & 6.32  & 6.35  & 6.81  \\
  $\sigma_{d}$  & 0.85   & 0.75    & 0.74   & 0.70  & 0.70  & 0.57  \\
 \hline
\end{tabular}
\label{table:all-algos-hot}
\end{center}
\end{table}

\begin{table}
\caption{\footnotesize{\bf Scalar metrics for $3K$-random HOT graphs generated using different techniques.} }
\begin{tabular}{|l|l|l|l|}
  \hline
  Metric &  $3K$-randomizing &  $3K$-targeting & Original \\
& rewiring & rewiring & HOT \\
  \hline
  $\bar{k}$    & 2.10  &  2.13   & 2.10   \\
   $r$         & -0.22  &  -0.23   & -0.22 \\
  $\bar{d}$    & 6.55  & 6.79    & 6.81   \\
  $\sigma_{d}$ & 0.84  & 0.72    & 0.57  \\
 \hline
\end{tabular}
\label{table:3k-algos-hot}
\end{table}

We first compare the different graph generation algorithms discussed in
Section~\ref{sec:dk-constructions}.
All the algorithms give consistent results,
except the stochastic approach, which suffers from the problems related to high statistical variance
discussed in Section~\ref{sec:dk-stochastic}. This conclusion immediately follows from
Figure~\ref{fig:comp-all-algos} and Tables~\ref{table:all-algos-hot} and \ref{table:3k-algos-hot}
showing graph metric values for the different $2K$ and $3K$~algorithms
described in Section~\ref{sec:dk-constructions}.

In our experience, we find that $dK$-randomizing rewiring is easiest to use.
However, it requires the original graph as input. If
only the target $dK$-distribution is
available and if~\mbox{$d \leq 2$}, we find the pseudograph algorithm most appropriate
in practice. We note that our $2K$ version results in fewer pseudograph ``badnesses'',
i.e., (self-)loops and small connected components~(CCs),
than PLRG~\cite{AiChLu00}, its commonly-known $1K$ counterpart.
This improvement is due to the additional constraints introduced by the $2K$ case.
For example, if there is only one node of high degree~$x$ and one node of
another high degree~$y$ in the original graph, then there can be only one
link of type~\mbox{$(x,y)$}. Our $2K$~modification of the pseudograph algorithm
must consequently produce exactly one link between these two $x$- and $y$-degree nodes,
whereas in the $1K$~case, the algorithm tends to create many such links.
Similarly, a $1K$ generator tends to produce many pairs of isolated
$1$-degree nodes connected to each another.
Since the original graph does not have such pairs, i.e., \mbox{(1,1)}-links,
our $2K$~generator, as opposed to~$1K$, does not form these small $2$-node CCs either.

While the pseudograph algorithm is a good $2K$-random graph
generator, we could not generalize it for~\mbox{$d \geq 3$} (see Section~\ref{sec:dk-pseudograph}). Therefore, to generate
$dK$-random graphs with $d \geq 3$ when an original graph is
unavailable, we use $dK$-targeting rewiring. We first bootstrap
the process by constructing $1K$-random graphs using the
pseudograph algorithm and then apply $2K$-targeting
$1K$-preserving rewiring to obtain $2K$-random graphs. To produce
$3K$-random graphs, we apply $3K$-targeting $2K$-preserving
rewiring to the $2K$-random graphs obtained at the previous step.

\subsection{Topology Comparisons}

We next test the convergence of our $dK$-series for the skitter and HOT graphs.
Since all $dK$-graph constructing algorithms yield consistent results, we selected
the simplest one, the $dK$-randomizing rewiring from Section~\ref{subsec:dk-rewiring},
to obtain $dK$-random graphs in this section.

The number of possible initial $dK$-randomizing rewirings is a good preliminary
indicator of the size of the $dK$-graph space.
We show these numbers for the HOT graph in Table~\ref{table:hot-initial-rewirings}.
If we discard rewirings leading to obvious isomorphic graphs,
cf.~Section~\ref{subsec:dk-rewiring},
then the number of possible initial rewirings is even smaller.

\begin{table}
\caption{\footnotesize{\bf Numbers of possible initial $dK$-randomizing rewirings for the HOT graph.} }
\begin{tabular}{|p{0.25in}|l|l|}
  \hline
  $d$ & Possible initial & Possible initial rewirings,\\
&  rewirings & ignoring obvious isomorphisms\\
  \hline
0 & 435,546,699 & - \\
1 & 477,905  & 440,355 \\
2 & 326,409  & 268,871 \\
3 & 146 & 44 \\
 \hline
\end{tabular}
\label{table:hot-initial-rewirings}
\end{table}

\begin{figure*}[tbh]
    \centerline{
        \subfigure[Distance distribution]
        {\includegraphics[width=2.1in]{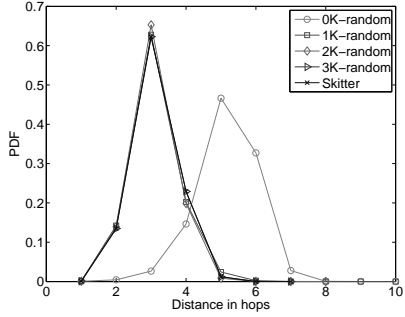}
        \label{fig:sk-dist} }
        \hfill
        \subfigure[Betweenness]
        {\includegraphics[width=2.1in]{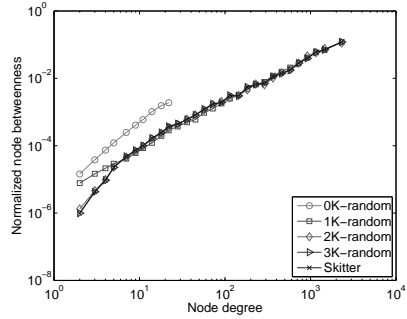}
        \label{fig:sk-bet} }
        \hfill
        \subfigure[Clustering]
        {\includegraphics[width=2.1in]{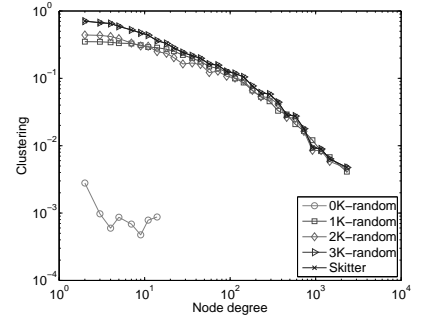}
        \label{fig:sk-clus} }
    }
    \caption{\footnotesize \bf Comparison of $dK$-random and skitter graphs.}
    \label{fig:sk-met}
\end{figure*}

\begin{table}
\caption{\footnotesize{\bf Comparing scalar metrics for $dK$-random and skitter graphs.}}
\begin{tabular}{|p{0.4in}|p{0.39in}|p{0.39in}|p{0.39in}|p{0.4in}|p{0.4in}|}
  \hline
  Metric & $0K$ & $1K$ & $2K$ & $3K$ & skitter \\
  \hline
  $\bar{k}$ & 6.31 & 6.34 & 6.29 & 6.29 & 6.29 \\
   $r$   & 0 & -0.24 & -0.24 & -0.24 & -0.24\\
  $\bar{C}$ & 0.001 & 0.25 & 0.29 & 0.46 & 0.46\\
  $\bar{d}$ & 5.17 & 3.11 & 3.08 & 3.09 & 3.12 \\
  $\sigma_{d}$ & 0.27 & 0.4 & 0.35 & 0.35 & 0.37  \\
    $\lambda_{1}$ & 0.2 & 0.03 & 0.15 & 0.1 & 0.1\\
    $\lambda_{n-1}$ & 1.8 & 1.97 & 1.85 &  1.9 & 1.9  \\
  \hline
\end{tabular}
\label{table:skitter}
\end{table}

We compare the skitter topology with its $dK$-random counterparts,
\mbox{$d=0,\ldots,3$},
in Table~\ref{table:skitter} and Figure~\ref{fig:sk-met}.
We report all the metrics calculated for the giant connected component~(GCCs).
Minor discrepancies between values of average degree~$\bar{k}$
and~$r$ result from GCC extractions. If we do not extract the
GCC, then~$\bar{k}$ is the
same as that of the original graph for all~\mbox{$d=0,\ldots,3$}, and~$r$ is exactly the same for~\mbox{$d>1$}.

We do not show degree distributions for brevity. However, degree distributions match
when considering the entire graph and are very similar for the GCCs for all~\mbox{$d>0$}.
When $d=0$, the degree distribution is binomial, as expected.

We see that all other metrics gradually converge to those in the original graph as~$d$ increases. A value of
$d=1$ provides a reasonably good description of the skitter topology, while $d=2$
matches all properties except clustering. The $3K$-random graphs are identical to the original
for all metrics we consider, including clustering.

\begin{figure*}[tbh]
  \begin{minipage}[t]{2in}
      \centerline{
          \includegraphics[width=2in]{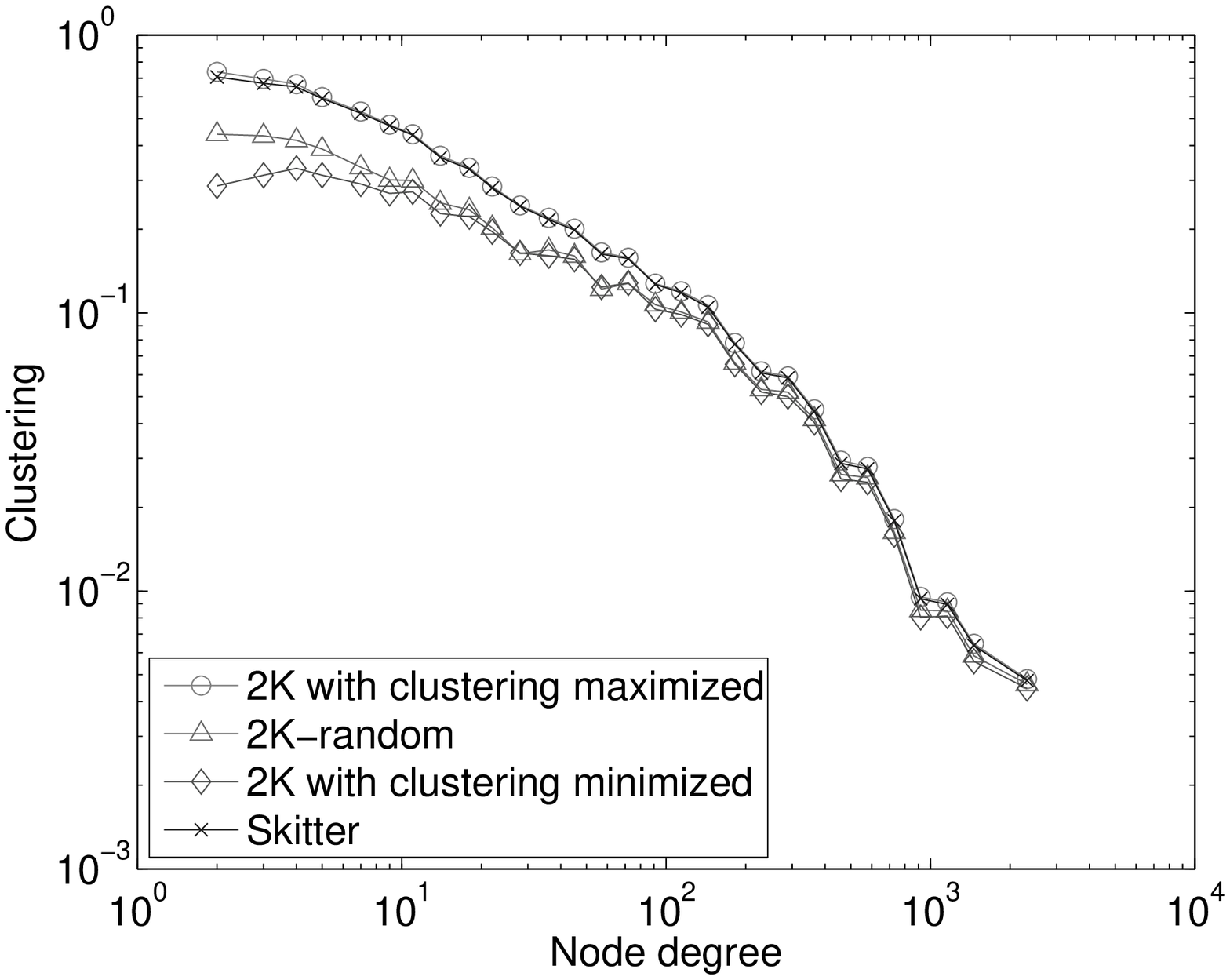}
      }
      \caption{\footnotesize \bf  Varying clustering in $2K$-graphs for skitter.
      }
      \label{fig:sk-vary-clus}
  \end{minipage}
  \hfill
  \begin{minipage}[t]{2in}
      \centerline{
          \includegraphics[width=2in]{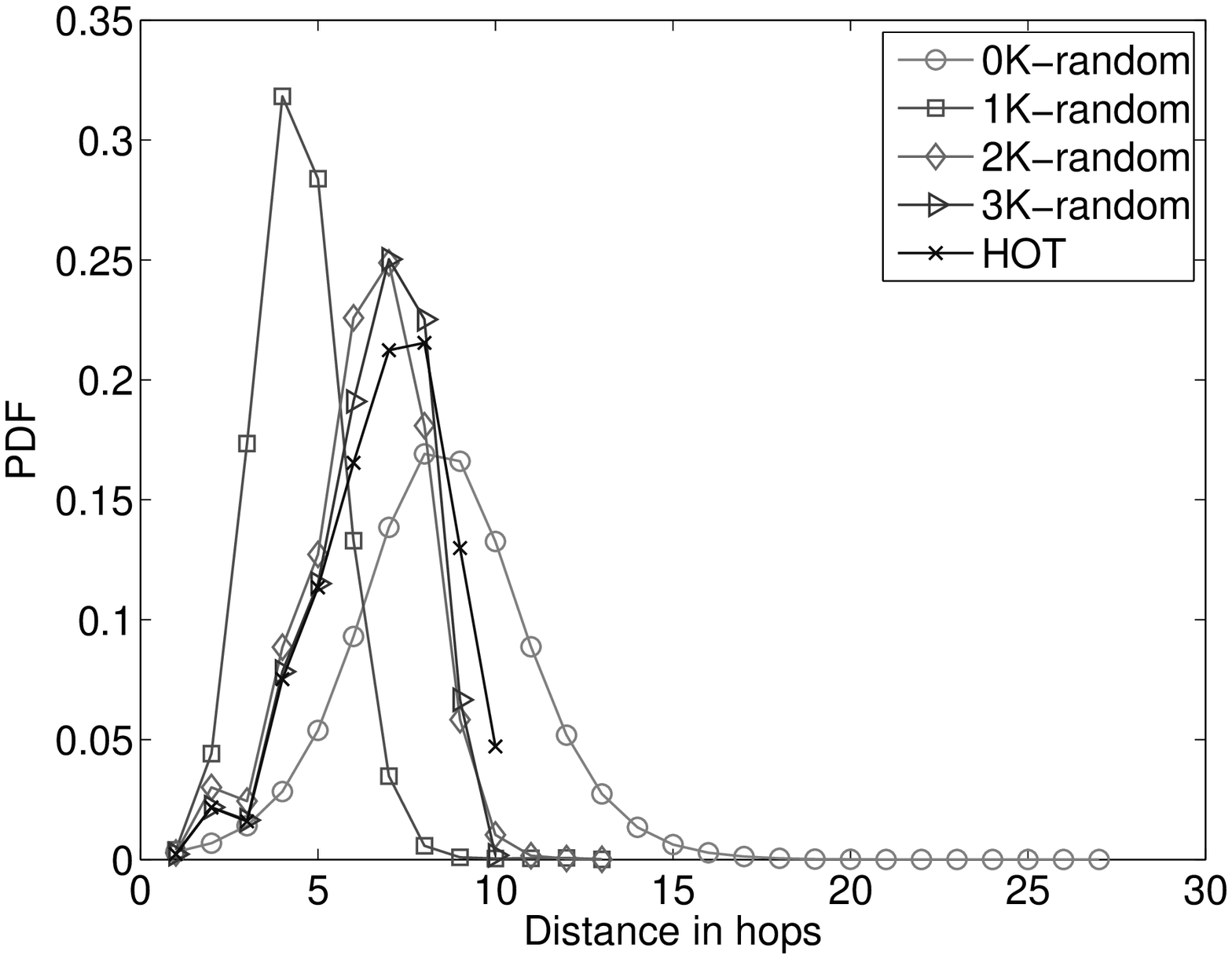}
      }
      \caption{\footnotesize \bf Distance distribution for $dK$-random and HOT graphs
      }
      \label{fig:hot-dist}
  \end{minipage}
\hfill
  \begin{minipage}[t]{2in}
      \centerline{
    \includegraphics[width=2in]{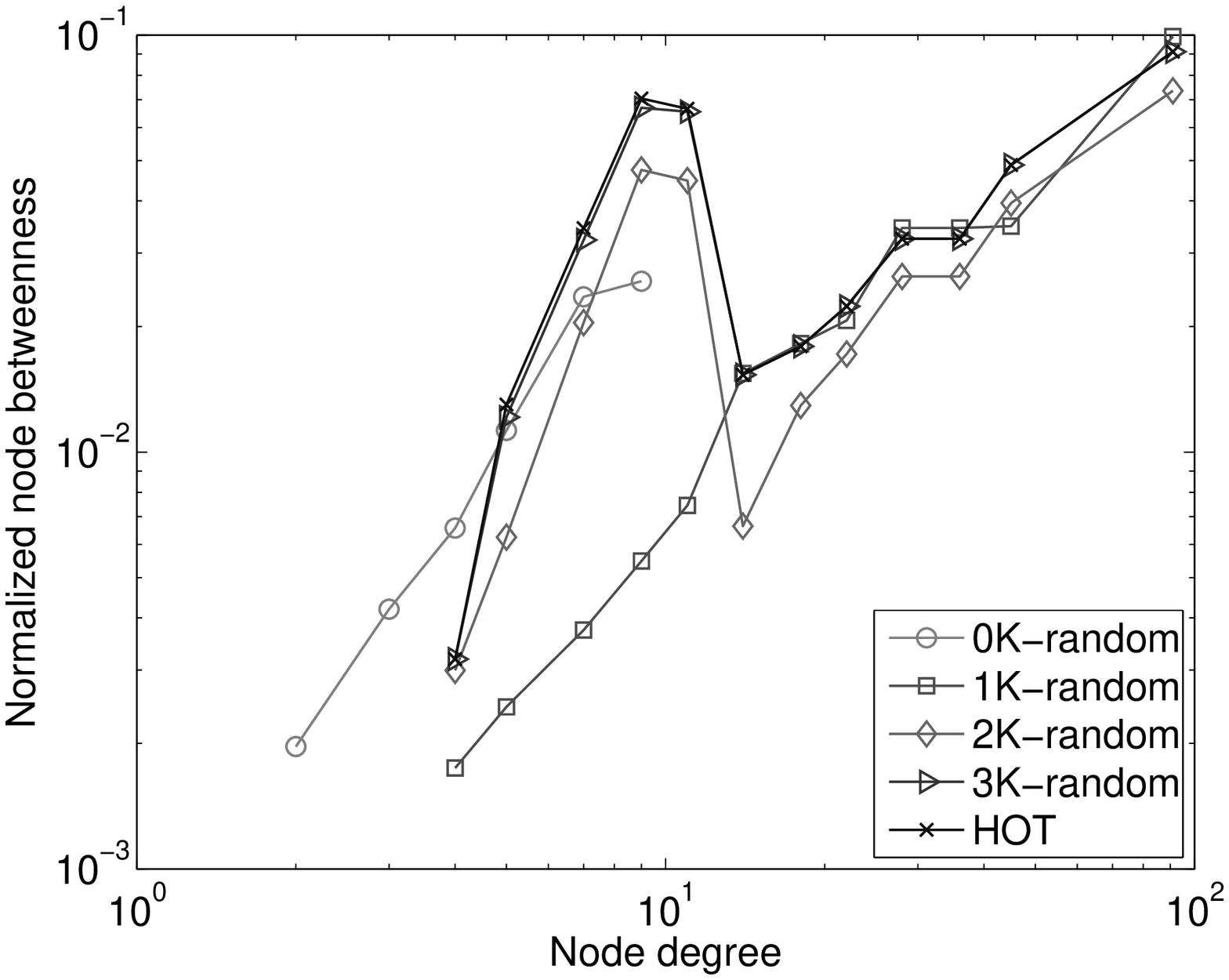}
    }
      \caption{\footnotesize \bf Betweenness for $dK$-random and HOT graphs
      }
      \label{fig:hot-bet}
 \end{minipage}
\end{figure*}

We perform the $2K$-space explorations described in Section~\ref{sec:dk-exploraitons}
to check if $d=2$ is indeed sufficiently constraining for the skitter topology. We observe small variations of
clustering~$\bar{C}$, second-order likelihood~$S_2$, and spectrum, shown
in Table~\ref{table:sk-exploration}
and Figure~\ref{fig:sk-vary-clus}. All other metrics do not change,
so we do not
show plots for them. We conclude that $d=2$, i.e., the joint degree distribution
provides a reasonably accurate description of observed AS-level topologies.

\begin{table}
\begin{center}
\caption{ \footnotesize \bf{Scalar metrics for $2K$-space explorations for skitter.}}
\begin{tabular}{|l|l|l|l|l|l|l|}
  \hline
  Metric & Min       & Max       & Min   & Max   & $2K$-  & Skit-\\
         & $\bar{C}$ & $\bar{C}$ & $S_2$ & $S_2$ & rand.  & ter  \\
  \hline
  $\bar{k}$ &6.29 & 6.29 & 6.29 & 6.29 & 6.29 & 6.29 \\
   $r$    & -0.24 & -0.24 & -0.24 & -0.24 & -0.24 & -0.24\\
  $\bar{C}$ & 0.21 & 0.47 & 0.4 & 0.4 & 0.29 & 0.46 \\
  $\bar{d}$ &  3.06 & 3.12 & 3.12 & 3.10 & 3.08 & 3.12 \\
  $\sigma_{d}$ & 0.33 & 0.38 & 0.37 & 0.36 & 0.35 & 0.37 \\
    $\lambda_{1}$   & 0.25  & 0.11    & 0.11 &0.1 & 0.15 & 0.1\\
    $\lambda_{n-1}$ & 1.75 & 1.89   & 1.89 &1.89  & 1.85 & 1.9 \\
    $S_{2}/S_{2}^{\max}$ & 0.988 &  0.961 & 0.955 & 1.000 & 0.986 & 0.958\\
   \hline
\end{tabular}
\label{table:sk-exploration}
\end{center}
\end{table}

The HOT topology is
more complex than AS-level topologies.
Earlier work argues that
this topology cannot be accurately modeled using degree
distributions alone~\cite{LiAlWiDo04}.
We therefore selected the HOT topology graph
as a difficult case for our approach.

A preliminary inspection of visualizations in Figure~\ref{fig:hot-cool-pic} indicates
that the $dK$-series converges at a reasonable rate even for the HOT graph.
The $0K$-random graph is a classical random graph and lacks high-degree nodes,
as expected.
The $1K$-random graph has all the high-degree nodes we desire,
but they are crowded toward the core,
a property absent in the HOT graph. The $2K$~constraints start pushing
the high-degree nodes away to the periphery, while the lower-degree nodes
migrate to the core, and the $2K$-random graph
begins to resemble the HOT graph.
The $3K$-random topology looks
remarkably similar to the HOT topology.

Of course, visual inspection of a small number of randomly generated graphs
is insufficient to demonstrate our ability to capture important metrics
of the HOT graph. Thus, we compute the different metric values for each of the $dK$-random
graph and compare them with the corresponding value
for the original HOT graph.
In Table~\ref{table:hot}
and Figures~\ref{fig:hot-dist} and~\ref{fig:hot-bet} we see that the $dK$-series converges more slowly for
HOT than for skitter. Note that we do not show clustering plots because
clustering is almost zero everywhere:
the HOT topology has very few cycles; it is almost a tree.
The $1K$-random graphs yield a poor approximation
of the original topology, in agreement with the main argument in~\cite{LiAlWiDo04}.
Both Figures~\ref{fig:hot-cool-pic}
and~\ref{fig:hot-bet} indicate that starting with \mbox{$d=2$}, low- but not high-degree
nodes form the core: betweenness is approximately as high for nodes of
degree $\sim 10$ as for high-degree nodes.
Consequently, the $2K$-random graphs provide a better approximation,
but not nearly as good as it was for skitter.\footnote{The speed of $dK$-series
convergence depends both on the structure and size of an original graph.
It must converge faster for smaller input graphs of similar structure. However, here we see that
the graph structure plays a more crucial role than its size. The $dK$-series
converges slower for HOT than for skitter, even though the former graph is an order
of magnitude smaller than the latter.}
However, the $3K$-random graphs match the original HOT topology
{\em almost exactly}.
We thus conclude that the $dK$-series captures the
essential characteristics of
even particularly difficult topologies, such as HOT,
by sufficiently increasing~$d$,
in this case to~3.
\begin{table}
\caption{\footnotesize {\bf Comparing scalar metrics for $dK$-random and HOT graphs.}}
\begin{tabular}{|p{0.4in}|p{0.4in}|p{0.4in}|p{0.4in}|p{0.4in}|p{0.4in}|}
  \hline
  Metric & $0K$ & $1K$ & $2K$ & $3K$ & HOT \\
  \hline
  $\bar{k}$ & 2.47& 2.59 & 2.18 & 2.10 & 2.10\\
   $r$    & -0.05 & -0.14 & -0.23 & -0.22 & -0.22 \\
  $\bar{C}$ & 0.002& 0.009& 0.001& 0 & 0 \\
  $\bar{d}$ & 8.48 & 4.41 & 6.32 & 6.55 & 6.81\\
  $\sigma_{d}$ & 1.23 & 0.72 & 0.71 & 0.84 & 0.57\\
    $\lambda_{1}$   & 0.01 & 0.034 & 0.005   & 0.004 & 0.004\\
    $\lambda_{n-1}$ & 1.989 & 1.967 & 1.996 & 1.997 & 1.997\\
  \hline
\end{tabular}
\label{table:hot}
\end{table}

\section{Discussion and Future Work}
\label{sec:discussion}
While we feel our approach to topology analysis holds significant
promise, a number of important avenues remain for further
investigation.  First, one must determine appropriate values
of~$d$ to carry out studies of interest.  Our experience to date
suggests that $d=2$ is sufficient to reproduce most metrics of
interest and that $d=3$ faithfully reproduces all metrics we are
aware of for Internet-like graphs.  It also appears likely that
$d=3$ will be sufficient for self-organized small-worlds in
general. This issue is particularly important because
the computational complexity of producing $dK$-graphs grows rapidly with~$d$.
Studies requiring large values of~$d$ may limit the practicality of
our approach.

In general, more complex topologies may necessitate developing
algorithms for generating $dK$-random graphs with high~$d$'s. We
needed higher~$d$ to describe the HOT topology as accurately as
the skitter topology. The intuition behind this observation is
that the HOT router-level topology is ``less random'' because it
results from targeted design and engineering.  The skitter
AS-level topology, on the other hand, is ``more random'' since
there is no single point of external human control over its shape
and evolution. It is a cumulative result of local decisions made
by individual ASes.

A second important question concerns the discrete nature of our
model. For instance, we are able to reproduce $1K$- and
$2K$-distributions but it is not meaningful to consider
reproducing $1.4K$-distributions. Consider a graph property~$X$
not captured by~$1K$ but successfully captured by~$2K$.  It could
turn out that the space of $2K$-random graphs over-constrains the
set of graphs reproducing~$X$.  That is, while $2K$-graphs do
successfully reproduce~$X$, there may be other graphs that also
match $X$ but are not $2K$-graphs.

Fundamental to our approach is that we seek to reproduce important
characteristics of a {\em given\/} network topology.  We cannot use
our methodology to discover laws governing the evolutionary growth of
a particular network. Rather, we are restricted to observing changes
in degree correlations in graphs over time, and then generating graphs
that match such degree correlations.  However, the goals of
reproducing important characteristics of a given set of graphs and
discovering laws governing their evolution are complementary and even
symmetric.

They are complementary because the $dK$-series can simplify the task
of validating particular evolutionary models.  Consider
the case where a researcher wishes to validate a model for Internet
evolution using historical connectivity information.  The
process would likely involve starting with an initial graph, e.g., reflecting
connectivity from five years ago, and generating a variety of larger
graphs, e.g., reflecting modern-day connectivity.  Of course, the
resulting graphs will not match known modern connectivity exactly.
Currently, validation would require showing that the
graph matches ``well enough'' for all known ad hoc graph properties.  Using the
$dK$-series however, it is sufficient to demonstrate that the
resulting graphs are $dK$-random for an appropriate value of $d$, i.e., constrained by the
$dK$-distribution of modern Internet graphs, with $d=3$ known to be
sufficient in this case.  As long as the resulting
graphs fall in the $dK$-random space, the nature of $dK$-randomness
explains any graph metric variation from ground truth.  This methodology
also addresses the issue of defining ``well enough'' above: $dK$-space
exploration can quantify the expected variation in ad hoc properties not
fully specified by a particular $dK$-distribution.

The two approaches are symmetric in that they both attempt to generate
graph models that accurately capture values of topology metrics
observed in real networks.  Both approaches have inherent tradeoffs
between accuracy and complexity.  Achieving higher accuracy with the
$dK$-series requires greater numbers of statistical {\em
constraints\/} with increasing~$d$. The number of these constraints is
upper-bound\-ed by~$n^d$ (the size of $dK$-distribution matrices) times
the number of possible simple connected $d$-sized
graphs~\cite{sloane-A001349}.\footnote{Although the upper bound of
possible constraints increases rapidly, sparsity of $dK$-distribution
matrices increases even faster. The result of this interplay is that
the number of non-zero elements of $dK$-distributions for any
given~$G$ increases with~$d$ first but then quickly decreases, and it
is surely~$1$ in the limit of~\mbox{$d=n$}, cf.~the example in
Section~\ref{sec:degcor}.}  Achieving higher accuracy with network
evolution modeling requires richer sets of system-specific
external parameters~\cite{ChJaWi06}.  Every such parameter
represents a {\em degree of freedom\/} in a model. By tuning larger
sets of external parameters, one can more closely match observed data.
It could be the case---which remains to be seen---that the number
of parameters needed for evolution modeling is smaller than the
number of constraints required by the $dK$-series to characterize
the modern Internet structure with the same degree of accuracy.
However, with the $dK$-series, the same set of constraints
applies to any networks, including social, biological, communication, etc.
With evolution modeling, one must develop a separate model for
each specific network.

Directions for future work all stem from the observation that the
$dK$-series is actually the simplest basis for statistical
analysis of correlations in complex networks. We can incorporate
any kind of technological constraints into our constructions.
In a router-level topology, for example, there is some dependency between the number of interfaces a
router has (node degree) and their average bandwidth
(betweenness/degree ratio)~\cite{LiAlWiDo04}. In light of
such observations, we can
simply adjust our rewiring algorithms (Section~\ref{subsec:dk-rewiring}) to not accept rewirings violating
this dependency. In other words, we can always consider ensembles
of $dK$-random graphs subject to various forms of external
constraints imposed by the specifics of a given network.

Another promising avenue for future work derives from the observation
that abstracting real networks as undirected graphs might lose too
much detail for certain tasks.
For example, in the AS-level topology case, the link types
can represent business AS relationships, e.g., customer-provider or
peering. For a router-level topology, we can label links with
bandwidth, latency, etc., and nodes with router manufacturer,
geographical information, etc. Keeping such {\em annotation\/}
information for nodes and links can also be useful for other types of
networks, e.g., biological, social, etc.
We can generalize the $dK$-series approach to study networks with
more sophisticated forms of annotations, in which case the
$dK$-series would describe correlations among different types of
nodes connected by different types of links within $d$-sized
geometries.  Given the level of constraint imposed by \mbox{$d=2$}
and \mbox{$d=3$} for our studied graphs and recognizing that
including annotations would introduce significant additional
constraints to the space of $dK$-graphs, we believe that
$2K$-random annotated graphs could provide appropriate
descriptions of observed networks in a variety of settings.

In this paper, all synthetic graphs' sizes equal to
a given graph's size. We are working on appropriate
strategies of rescaling the $dK$-distributions to arbitrary graph sizes.

\section{Conclusions}
\label{sec:conclusion}
Over the years, a number of important graph metrics have been proposed
to compare how closely the structure of two arbitrary graphs match and
to predict the behavior of topologies with certain metric values.
Such metrics are employed by networking researchers involved in
topology construction and analysis, and by those interested in protocol and
distributed system performance.
Unfortunately,
there is limited understanding of which metrics are appropriate for a
given setting and, for most proposed metrics, there are no known
algorithms for generating graphs that reproduce the target property.

This paper defines a series of graph structural properties to both
systematically characterize arbitrary graphs and to generate random graphs
that match specified characteristics of the original.
The $dK$-distribution is a collection of distributions describing
the correlations of degrees of $d$~connected nodes.
The properties~$\mathcal{P}_d$, \mbox{$d=0,\ldots,n$}, comprise
the $dK$-series. A random graph is said to have property~$\mathcal{P}_d$
if its $dK$-distribution has the same form as in a given graph~$G$.
By increasing the value of~$d$ in the series, it is possible to capture
more complex properties of~$G$ and, in the limit, a sufficiently large
value of~$d$ yields complete information about $G$'s structure.

We find interesting tradeoffs in choosing
the appropriate value of~$d$ to compare two graphs or to generate
random graphs with property~$\mathcal{P}_d$.  As we increase~$d$,
the set of randomly generated graphs having
property~$\mathcal{P}_d$ becomes increasingly constrained and the
resulting graphs are increasingly likely to reproduce a variety of
metrics of interest.  At the same time, the algorithmic complexity
associated with generating the graphs grows sharply.  Thus, we
present a methodology where practitioners choose the smallest~$d$ that
captures essential graph characteristics for their study.  For the
graphs that we consider, including comparatively complex Internet
AS- and router-level topologies, we find that $d=2$ is sufficient for most cases
and $d=3$
captures all graph
properties proposed in the literature known to us.

In this paper, we present the first algorithms for constructing
random graphs having properties~$\mathcal{P}_2$ and~$\mathcal{P}_3$,
and sketch an approach for extending the algorithms to arbitrary~$d$.
We are also releasing the source code for our analysis
tools to measure an input graph's $dK$-distribution and our
generator able to produce random graphs possessing
properties~$\mathcal{P}_d$ for~\mbox{$d<4$}.

We hope that our methodology will provide
a more rigorous and consistent method of comparing topology
graphs and enable protocol and application researchers to test
system behavior under a suite of randomly generated yet appropriately
constrained and realistic network topologies.

\section*{Acknowledgements}
We would like to thank Walter Willinger and Lun Li for their HOT
topology data; Bradley Huffaker, David Moore, Marina Fomenkov, and
kc claffy for their contributions at different stages of this work;
and anonymous reviewers for their comments that helped to improve
the final version of this manuscript.
Support for this work was provided by NSF CNS-0434996 and Center for
Networked Systems~(CNS) at UCSD.

\bibliographystyle{abbrv}
\vspace*{0.05in}
\scriptsize
\bibliography{paper}

\end{document}